\documentclass[prd, twocolumn, nofootinbib, superscriptaddress]{revtex4-2}

\usepackage{aas_macros}
\usepackage{graphicx}
\usepackage[caption=false]{subfig}
\usepackage{siunitx}
\usepackage{mathrsfs}
\usepackage{amsmath,amssymb}
\usepackage{bm}
\usepackage{braket}
\usepackage{listings}
\usepackage{cases}
\usepackage{here}
\usepackage{comment}
\usepackage{soul}
\usepackage{cancel}
\usepackage{cases}
\usepackage[utf8]{inputenc}
\usepackage{url}
\usepackage{longtable}
\usepackage[normalem]{ulem}
\usepackage{xspace}
\usepackage[colorlinks=true
,urlcolor=blue
,anchorcolor=blue
,citecolor=blue
,filecolor=blue
,linkcolor=blue
,menucolor=blue
,linktocpage=true
,pdfproducer=medialab
,pdfa=true
]{hyperref}
\usepackage{animate}

\begin{document}


\title{Analytic model for the statistics of ultra-high magnification events}

\author{Hiroki Kawai}
\email{hiroki.kawai@phys.s.u-tokyo.ac.jp}
\affiliation{Department of Physics, The University of Tokyo, Bunkyo, Tokyo 113-0033, Japan}
\affiliation{Center for Frontier Science, Chiba University, 1-33 Yayoicho, Inage, Chiba 263-8522, Japan}
\affiliation{INAF – Osservatorio Astronomico di Bologna, via Ranzani 1, 40127 Bologna, Italy}

\author{Masamune Oguri}
\affiliation{Center for Frontier Science, Chiba University, 1-33 Yayoicho, Inage, Chiba 263-8522, Japan}
\affiliation{Department of Physics, Graduate School of Science, Chiba University, 1-33 Yayoicho, Inage, Chiba 263-8522, Japan}

\date{\today}

\begin{abstract}
Highly magnified individual stars such as Icarus and Earendel have recently been observed near critical curves of galaxy clusters with Hubble Space Telescope (HST) observations.
These stars are estimated to be magnified with a factor of more than a few thousands.
In addition to the smooth mass distribution in the macro-lens model, the distribution of microlenses originating from, for instance, intracluster stars affects the event rate and the peak magnification significantly.
We propose an analytic model of the high-magnification tail of the probability distribution function (PDF) in which the probability is assumed to be proportional to the number of independent microlens critical curves.
Our model can explain the parameter dependence of the PDF on the mass fraction of the microlenses and the background magnification seen in ray-tracing simulations.
The effect of a finite source size is also studied to derive a fitting formula for the suppression factor.
For an application of our model, we calculate the event rate of the Icarus-like system and the probability distribution of observed positions of such a system, showing good agreement with the HST observations. Our model predicts a complicated dependence of the probability distribution of observed positions of highly magnified events on the magnification threshold.
\end{abstract}


\maketitle

\section{Introduction} \label{sec:intro}
With the improvement in the performance of observations, it has become possible to find more gravitational lensing events.
Especially, highly magnified individual stars have recently been found near critical curves of galaxy clusters.
The first example is Icarus at redshift $z=1.49$ observed in the two-year survey by the Hubble Space Telescope (HST), which was magnified with a factor of more than two thousands due to the MACS J1149 cluster at redshift $z=0.544$ \citep{2018NatAs...2..334K}.
The peak magnification continues for less than two weeks, from which it is concluded that lensing is mainly caused by microlenses in the cluster such as stars that are responsible for the intracluster light (ICL).
A high-redshift star at $z=6.7$ called Earendel has also been reported by the HST observation \citep{2022Natur.603..815W}.
Earendel is located on the critical curve of the galaxy cluster WHL0137–08 at redshift $z=0.566$ with an estimated magnification of more than a thousand.
The peak magnification continues over 3.5 years.
These phenomena are explained by the corrugated band of microlenses near the critical curve (see \cite{2024arXiv240408094W} for a recent review).
Although both Icarus and Earendel are highly magnified near the critical curves of the galaxy clusters in the presence of microlenses, the locations and durations of the peak magnifications differ.
The observed number of such high magnification events is rapidly increasing ({\it{e.g.,}} \citep{2023A&A...679A..31D, 2024arXiv240408045F}), which opens the possibility of studying these events statistically.

These highly magnified stars should be observed near critical curves of lensing objects, whose shapes depend on the mass distributions within these objects. 
Critical curves can broadly be categorized into two types.
The first type arises from the overall density distribution of lens objects, forming macro-critical curves typically about a few tens of arcsec in size in the case of galaxy clusters.
The shapes of the macro-critical curves are perturbed by the presence of subhalos \citep{2024PhRvD.109h3517A}.
Cold dark matter subhalos with mass $M_{\rm h} = 10^{6}$-$10^{8}\ M_{\odot}$ produce the distortions of about 10 milli-arcssec \citep{2018ApJ...867...24D}.
The second type consists of micro-critical curves within lens objects, generated by microlenses such as stars and black holes, typically much smaller than 0.1 arcsec in size.
In the case of galaxy clusters, ICL stars contribute to the microlenses.
Due to the presence of these small-scale structures in the lens objects, the observed signatures of highly magnified stars are affected ({\it{e.g.,}} \citep{2023arXiv230406064W}).

The presence of microlenses yields frequent peak magnifications and many micro-images due to their micro-critical curves.
\citet{2017ApJ...850...49V} analytically derive the characteristic properties such as 
the width of the corrugated network near the macro-critical curve, the frequency of caustic-crossing events, and the peak magnifications. 
Using their estimation, they constrain the parameter region of massive compact halo objects as dark matter.
\citet{2018ApJ...857...25D} conducted detailed ray-tracing simulations of the corrugated network near the macro-critical curve.
\citet{2018PhRvD..97b3518O} also study the effect of microlenses near the critical curves and calculate the event rate of Icarus with the help of the software called {\sc Glafic} \citep{2010PASJ...62.1017O}.
\citet{2020AJ....159...49D} consider the presence of axion minihalos in galaxy clusters and show that such minihalos produce surface density fluctuations and irregular light curves.
\citet{2021arXiv210412009D} extend an analytic work by \citet{1986ApJ...306....2K} and analytically study the random microlensing field, deriving an analytic expression for the average and its dispersion of the magnification.
\citet{2014ApJS..211...16V} and \citet{2024A&A...687A..81P} conduct inverse ray shooting simulations to obtain probability distribution functions (PDFs) with different parameters.
\citet{2024A&A...687A..81P} obtain the empirical formulae to describe PDFs.
While these studies significantly advance our understanding of the statistics of highly magnified events near critical curves of galaxy clusters, a missing piece is a physical model of the high magnification tail of the PDF that is applicable for a wide range of parameters, which motivates our study.

In this paper, we propose an analytic model of the high-magnification tail of the PDF that is based on the number of independent micro-critical curves.
We first obtain the PDF in the case of point sources with the help of simulation {\sc CCtrain}, which is a modified version of {\sc Glafic} \citep{2010PASJ...62.1017O} to calculate the caustic-crossing events.
We find that the combination of the surface mass density of microlenses and the background (average) magnification is an important quantity for describing the PDF in all parameter regions.
We then consider the case with finite source sizes, which limit the maximum magnification and suppress the high-magnification tail of the PDF.
With the help of the simulation of GPU-Enabled High-Resolution cosmological MicroLensing parameter survey ({\sc Gerlumph}) \citep{2014ApJS..211...16V}, we find that the suppression of the PDF is well described with the sigmoid function.
Our model agrees well with these simulation data, including the parameter dependence on the background magnification, the mass fraction of microlenses, and the effect of a finite source size.
For the application of our model, we focus on the Icarus-like system to calculate the event rate of highly magnified stars and the probability distribution of their observed positions.
Here we also consider the PDF around the averaged magnification modeled by \citet{2021arXiv210412009D} for better predictions.
Since Icarus continuing for roughly two weeks is found in the two-year survey of the HST, the observed mean number of Icarus-like events in each snapshot can be approximated as $1/52 \simeq 0.019$, which we confront with our theoretical model prediction.

In Sec.~\ref{sec:review}, we review the general theory on the point mass lenses that we use in this paper.
We show our analytic model for the high-magnification tail of the PDF in Sec.~\ref{sec:analytical_pdf}, where we introduce the notion of the independent critical curves.
In Sec.~\ref{sec:comparison_with_simulation}, we show the validity of our model by comparing it with simulations.
As a possible application of our model, we calculate the event rate of highly magnified stars and the probability distribution of their observed locations in the Icarus-like system, which we show in Sec.~\ref{sec:application}.
We finally show the summary and discussions in Sec.~\ref{sec:discussion_conclusion}.

\section{General theory} \label{sec:review}
In this section, we review the general theory of point mass lenses, which we apply to the model proposed in this paper.

\subsection{Magnification near critical curves} \label{subsec:mu_near_cc}
The gravitational lens equation maps the coordinates on the lens plane $\boldsymbol{\theta} = (\theta_{1}, \theta_{2})$ to the coordinates on the source plane $\boldsymbol{\beta} = (\beta_{1}, \beta_{2})$ as
\begin{equation}
    \boldsymbol{\beta} = \boldsymbol{\theta} - \boldsymbol{\alpha}(\boldsymbol{\theta}),
\end{equation}
where $\boldsymbol{\alpha}(\boldsymbol{\theta})$ is a deflection angle.
The deflection angle can be calculated from the lens potential $\psi(\boldsymbol{\theta})$ as
\begin{equation}
    \boldsymbol{\alpha}(\boldsymbol{\theta}) = \nabla  \psi(\boldsymbol{\theta}) \label{defang_pot},
\end{equation}
and is related to the mass distribution of the lens objects.

Since we focus on lensing properties near the critical curve in this paper, we choose the origins of the lens and source planes as points on a critical curve and a caustic, respectively \citep{2024PhRvD.109h3517A, 2024arXiv240408094W}.
In this case, the lens potential can be expressed as (see derivation in App.~\ref{app:lens_potential})
\begin{equation}
    \psi(\boldsymbol{\theta}) = \frac{1}{2} \{\kappa_{0}(\theta_{1}^{2} + \theta_{2}^{2}) + (1-\kappa_{0})(\theta_{1}^{2} - \theta_{2}^{2}) \} - \frac{\epsilon}{6}\theta_{1}^{3}, \label{lens_pot_near_cc}
\end{equation}
where $\kappa_{0}$ is the convergence on the origin and $\epsilon$ is the minus of the third derivative of the potential for $\theta_{1}$ at the origin and is proportional to the curvature of the critical curve.
From Eq.~\eqref{defang_pot}, the deflection angle is computed as
\begin{equation}
    \boldsymbol{\alpha}(\boldsymbol{\theta}) = 
    \begin{pmatrix}
        \theta_{1} - \frac{\epsilon}{2} \theta_{1}^{2} \\
        (2\kappa_{0} - 1) \theta_{2}
    \end{pmatrix}, \label{def_ang_near_cc}
\end{equation}
and the lens equations are
\begin{eqnarray}
    &&\beta_{1} = \frac{\epsilon}{2}\theta_{1}^{2}, \label{lens_eq1}\\
    &&\beta_{2} = 2(1-\kappa_{0})\theta_{2} \label{lens_eq2}.
\end{eqnarray}
From Eqs.~\eqref{lens_eq1} and \eqref{lens_eq2}, the Jacobian lens matrix can be calculated as 
\begin{equation}
    A(\boldsymbol{\theta}) = \frac{\partial \boldsymbol{\beta}}{\partial \boldsymbol{\theta}} = 
    \begin{pmatrix}
       \epsilon \theta_{1} & 0\\
        0 & 2(1-\kappa_{0})
    \end{pmatrix}.
\end{equation}
The convergence and the shear are calculated from the lens potential as
\begin{eqnarray}
    &&\kappa(\boldsymbol{\theta}) = \frac{1}{2} (\psi_{,11} + \psi_{,22}) =  \kappa_{0} - \frac{\epsilon}{2} \theta_{1}, \\
    && \gamma_{1}(\boldsymbol{\theta}) = \frac{1}{2} (\psi_{,11} - \psi_{,22}) = 1 -\kappa_{0} - \frac{\epsilon}{2} \theta_{1},\\
    && \gamma_{2}(\boldsymbol{\theta}) = \psi_{,12} = 0 ,\\
    && \gamma(\boldsymbol{\theta}) = \sqrt{\gamma_{1}(\boldsymbol{\theta})^{2} + \gamma_{2}(\boldsymbol{\theta})^{2}} = 1 -\kappa_{0} - \frac{\epsilon}{2} \theta_{1}.
\end{eqnarray}
Here the subscription of $,1$ and $,2$ represents the derivative with respect to $\theta_{1}$ and $\theta_{2}$, respectively.
The magnification is the inverse of the determinant of the Jacobian matrix and is given by
\begin{equation}
    \mu(\boldsymbol{\theta}) = \frac{1}{|\det A(\boldsymbol{\theta})|} = \frac{1}{2\epsilon (1-\kappa_{0})} \cdot \frac{1}{\theta_{1}} . \label{mu_near_crit}
\end{equation}
Equation~\eqref{mu_near_crit} shows that the critical line is the straight line on the lens plane, $\theta_{1} = 0$, and magnification is inversely proportional to the distance from the critical curve.
Using Eq.~\eqref{lens_eq1}, we can obtain the magnification on the source plane as
\begin{equation}
    \mu(\boldsymbol{\beta}) = \frac{1}{\sqrt{2\epsilon} (1-\kappa_{0})} \cdot \frac{1}{\sqrt{\beta_{1}}}.\label{mu_near_caustic}
\end{equation}
To obtain this relation, we multiply a factor of two since two images with the same magnification emerge symmetrically on both sides of the critical curve when the source is located outside the caustic.
The caustic is the straight line on the source plane, $\beta_{1} = 0$, and the dependence of the distance on the magnification is $\mu \propto \beta^{-1/2}$.

When the source has a finite source size $\sigma_{\rm W}$, the magnification is averaged as 
\begin{eqnarray}
    \mu(\boldsymbol{\beta}) = && \frac{1}{\sqrt{2\epsilon}(1-\kappa_{0})}\frac{1}{\sqrt{2\pi \sigma_{\rm W}^{2}}} \nonumber \\
    && \times  \int_{0}^{\infty} d\beta'_{1} \frac{1}{\sqrt{\beta'_{1}}} \exp\left[-\frac{(\beta'_{1}-\beta_{1})^{2}}{2\sigma_{\rm W}^{2}}\right]. \label{mu_beta_source_eff}
\end{eqnarray}
Here we consider a Gaussian surface brightness profile.
In this case, the peak magnification is given by
\begin{equation}
    \mu_{\rm max} \simeq \frac{1}{\sqrt{2\epsilon} (1-\kappa_{0})} \cdot \frac{1}{\sqrt{\sigma_{\rm W}}}.\label{max_mu_eps}
\end{equation}

Next, we consider the area on the lens plane where the magnification is larger than $\mu$.
This can be obtained by integrating along the critical curve using Eq.~\eqref{mu_near_crit} 
\begin{equation}
    a_{\rm l}(>\mu) \simeq \oint d\lambda \frac{dl}{d\lambda} \frac{1}{2\epsilon (1-\kappa_{0})\mu} \propto \frac{1}{\mu},
\end{equation}
where $\lambda$ is the parameter to describe the critical curve.
The area where the magnification is between $\mu$ and $\mu+d\mu$ is
\begin{equation}
    \left|\frac{da_{\rm l}(>\mu)}{d\mu}\right|d\mu \propto \frac{1}{\mu^{2}}d\mu.
\end{equation}
The cross-section is defined by the area on the source plane where the magnification becomes between $\mu$ and $\mu+d\mu$, which can be obtained by mapping the above area onto the source plane
\begin{equation}
    \left|\frac{da_{\rm s}(>\mu)}{d\mu}\right|d\mu = \frac{2}{\mu}\left|\frac{da_{\rm l}(>\mu)}{d\mu}\right|d\mu \propto \frac{1}{\mu^{3}}d\mu,
\end{equation}
where a factor of two emerges in the middle term since the images are found on both sides of the critical curves.
Thus we can obtain the area on the source plane where the magnification is larger than $\mu$ as
\begin{equation}
    a_{\rm s}(>\mu) \propto \frac{1}{\mu^{2}}.
\end{equation}
The above relations indicate that the high-magnification tail of the PDF behaves as
\begin{equation}
    \frac{dp}{d\mu} \propto \mu^{-3},
\end{equation}
or equivalently,
\begin{equation}
    \frac{dp}{d\log_{10} \mu} \propto \mu^{-2} \label{pdf_tail_mu2}.
\end{equation}

\subsection{Microlens in a smooth mass distribution} \label{subsec:microlens_in_smooth}
In this subsection, we show that the size of the critical curve becomes larger when a smooth mass distribution exists in the background \citep{2018PhRvD..97b3518O}.

We first briefly review gravitational lensing by a point mass lens without the background.
The lens equation is expressed by
\begin{equation}
    \boldsymbol{\beta} = \boldsymbol{\theta} - \frac{\theta_{\rm Ein}^{2}}{\theta^{2}}\boldsymbol{\theta}, \label{lens_eq_point_wo_bg}
\end{equation}
with $\theta = |\boldsymbol{\theta}|$.
Note that now we set the origin of the lens plane at the position of the point mass lens.
The Einstein radius $\theta_{\rm Ein}$ is expressed by
\begin{equation}
    \theta_{\rm Ein} = \sqrt{\frac{4GM}{c^{2}}\frac{D_{\rm ls}}{D_{\rm os}D_{\rm ol}}}, \label{einstein_radius}
\end{equation}
where $G$ is the Newton constant, $D_{\rm ls}, D_{\rm os}$, and $D_{\rm ol}$ are the angular distances between lens and source, observer and source, and observer and lens, respectively.
From the lens equation, we can obtain the magnification as 
\begin{equation}
    \mu(\theta) = \frac{1}{1-\left(\frac{\theta_{\rm Ein}}{\theta}\right)^{4}}. \label{mag_point_wo_bg}
\end{equation}

Next, we consider a point mass lens embedded in the constant convergence $\kappa_{\rm B}$ and (absolute) shear $\gamma_{\rm B}$ as the background.
In this scenario, the lens equation has an additional term and is modified from Eq.~\eqref{lens_eq_point_wo_bg} as
\begin{eqnarray}
    \beta_{1} = \frac{\theta_{1}}{\mu_{\rm t,B}} - \frac{\theta_{\rm Ein}^{2}\theta_{1}}{\theta^{2}}, \label{lens_eq1_point_w_bg} \\
    \beta_{2} = \frac{\theta_{2}}{\mu_{\rm r,B}} - \frac{\theta_{\rm Ein}^{2}\theta_{2}}{\theta^{2}} \label{lens_eq2_point_w_bg},
\end{eqnarray}
where we introduce the tangential and radial magnification by the background smooth component
\begin{eqnarray}
    && \mu_{\rm t,B} = (1-\kappa_{\rm B}-\gamma_{\rm B})^{-1}, \\
    && \mu_{\rm r,B} = (1-\kappa_{\rm B}+\gamma_{\rm B})^{-1}.
\end{eqnarray}
Note that the total magnification is $\mu_{\rm B} = \mu_{\rm t,B} \mu_{\rm r,B}$.
Since we focus on the region near the tangential critical curve, $\mu_{\rm t,B}$ is much larger than $\mu_{\rm r,B} \simeq 1$.
The Jacobian matrix is 
\begin{equation}
    A(\boldsymbol{\theta}) =
    \begin{pmatrix}
       \mu_{\rm t,B}^{-1}+\frac{\theta_{\rm E}^{2}}{\theta^{2}}\cos 2\phi & \frac{\theta_{\rm E}^{2}}{\theta^{2}} \sin 2\phi\\
        \frac{\theta_{\rm E}^{2}}{\theta^{2}} \sin 2\phi & \mu_{\rm r,B}^{-1}-\frac{\theta_{\rm E}^{2}}{\theta^{2}}\cos 2\phi
    \end{pmatrix},
\end{equation}
where $\phi$ is a polar angle, $\phi = \arctan (\theta_{2}/\theta_{1})$.
The determinant of the Jacobian matrix is
\begin{equation}
    \det A = \mu_{\rm t,B}^{-1} \mu_{\rm r,B}^{-1} + (\mu_{\rm r,B}^{-1} -\mu_{\rm t,B}^{-1}) \frac{\theta_{\rm Ein}^{2}}{\theta^{2}}\cos 2\phi - \frac{\theta_{\rm Ein}^{4}}{\theta^{4}}.
\end{equation}
In the direction of $\phi=\frac{\pi}{2}$, we can express the magnification as follows by using the above relation, 
\begin{eqnarray}
    \mu(\theta) &=& \mu_{\rm t,B} \mu_{\rm r,B} \frac{1}{1-(\mu_{\rm t,B} -\mu_{\rm r,B})\frac{\theta_{\rm E}^{2}}{\theta^{2}} - \mu_{\rm t,B} \mu_{\rm r,B} \frac{\theta_{\rm Ein}^{4}}{\theta^{4}}}\nonumber \\
    &\simeq& \mu_{\rm t,B} \mu_{\rm r,B}\frac{1}{1-\left(\frac{\mu_{\rm t,B}\theta_{\rm Ein}^{2}}{\theta^{2}}\right)^{2}}. \label{mag_point_w_bg}
\end{eqnarray}
By comparing Eqs.~\eqref{mag_point_wo_bg} and \eqref{mag_point_w_bg}, we can see that the size of the Einstein radius becomes larger by a factor of $\sqrt{\mu_{\rm t,B}}$ due to the background smooth component.
This can be understood from the fact that the area of the tangential critical curve expands approximately by a factor of $\mu_{\rm B} \simeq \mu_{\rm t,B}$.
Note that the actual shape of the critical curve is gourd-shaped as already shown by \citet{2018PhRvD..97b3518O}.

The behavior of the magnification near the critical curve and caustic can be estimated as follows.
By expanding the magnification obtained in Eq.~\eqref{mag_point_w_bg} around the position of the critical curve, $\theta = \sqrt{\mu_{\rm t,B}} \theta_{\rm Ein} + \Delta \theta$, we can first obtain the magnification as a function of the distance to the critical curve on the lens plane as
\begin{equation}
    \mu(\Delta \theta) = \frac{1}{2} \mu_{\rm t,B} \mu_{\rm r,B} \left(\frac{\sqrt{\mu_{\rm t,B}} \theta_{\rm Ein}}{\Delta \theta}\right).\label{mag_near_crit_w_bg}
\end{equation}
We can check that the dependence is the same as in Eq.~\eqref{mu_near_crit}.
Next, we consider the magnification near the caustic on the source plane.
Since the caustic has an asteroid-like shape as already shown by \citet{2018PhRvD..97b3518O} and the direction $\phi=\pi/2$ corresponds to the cusp, where the caustic-crossing occurs rarely, we focus on the direction of $\phi \sim \pi/2$ (not equal).
In this direction, the normal vector points of the caustic approximately toward the $\beta_{1}$ axis.
Let us consider the point on the source plane $(\beta_{1},\beta_{2}) = (\Delta\beta_{1}, \theta_{\rm Ein} \sqrt{\mu_{\rm t,B}}/\mu_{\rm r,B})$, where $\Delta\beta_{1}$ is the distance to the caustic and the second component corresponds to the size of the caustic toward the $\beta_{2}$ axis.
The corresponding point on the lens plane is $(\theta_{1}, \theta_{2}) = (\Delta\theta_{1}, \sqrt{\mu_{\rm t,B}}\theta_{\rm Ein})$, where $\Delta\theta_{1}$ satisfies 
\begin{equation}
    \Delta\theta_{1} = \sqrt{\frac{1}{2} \mu_{\rm t,B}^{\frac{3}{2}} \theta_{\rm Ein} \Delta \beta_{1}}. \label{deltatheta_delta_beta}
\end{equation}
Assuming that the distance to the critical curve $\Delta\theta$ can be estimated by $\Delta\theta_{1}$, we can obtain the magnification near the caustic by substituting Eq.~\eqref{deltatheta_delta_beta} into Eq.~\eqref{mag_near_crit_w_bg}
\begin{equation}
    \mu(\Delta \beta) = \frac{1}{\sqrt{2}} \mu_{\rm t,B} \mu_{\rm r,B} \sqrt{\frac{\theta_{\rm Ein}}{\sqrt{\mu_{\rm t,B}}\Delta \beta}}. \label{mag_near_caustic_w_bg}
\end{equation}
Here we change the notation $\Delta \beta_{1}$ to the general case $\Delta \beta$.
The behavior of the magnification is consistent with Eq.~\eqref{mu_near_caustic}.
While this approximation might not be fully accurate, Eq.~\eqref{mag_near_caustic_w_bg} is consistent with the relation numerically derived in \citet{2018PhRvD..97b3518O}.

\subsection{Randomly distributed microlenses in a smooth mass distribution} \label{subsec:review_random_ml}
In the previous subsection, we consider a single microlens in the smooth mass distribution.
In this subsection, we distribute microlenses randomly with many realizations and obtain the (ensemble) average of the total magnification and its variance.
Note that the total magnification refers to the sum of magnifications of micro-multiple images produced by microlenses.

Consider again the situation where the background convergence and shear are given by $\kappa_{\rm B}$ and $\gamma_{\rm B}$, respectively, and distribute microlenses with the same masses, {\it i.e.} the same Einstein radii, randomly.
The average convergence of microlenses is given by $\kappa_{\star}$.
\citet{2021arXiv210412009D} analytically show that the average magnification $\mu_{\rm av}= \langle \mu_{\rm tot} \rangle$ is written as
\begin{equation}
    \mu_{\rm av}(\boldsymbol{\beta}) = \frac{1}{2\pi (\sigma_{\rm ml}^{2} + \sigma_{\rm W}^{2})} \int d^{2}\theta \exp \left[ -\frac{|\boldsymbol{\theta} - \boldsymbol{\beta} - \boldsymbol{\alpha}_{\rm B}|^{2}}{2(\sigma_{\rm ml}^{2} + \sigma_{\rm W}^{2})} \right], \label{mu_av_general}
\end{equation}
where $\boldsymbol{\alpha}_{\rm B}$ is the deflection angle due to the background smooth mass distribution, $\sigma_{\rm ml}^{2}$ represents the variance of the random deflection angle by microlenses
\begin{equation}
    \sigma_{\rm ml}^{2}(R_{\star},l_{\star}) = \kappa_{\star} \theta_{\rm Ein}^{2} \left(1-\gamma_{\rm E} + \ln \frac{2R_{\star}}{\theta_{\rm Ein}^{2}l_{\star}}\right),
\end{equation}
where $\gamma_{\rm E}$ is the Euler–Mascheroni constant, and $R_{\star}$ and $l_{\star}$ are given by 
\begin{eqnarray}
    && R_{\star} = \mu_{\rm B} \sigma_{\rm eff}, \label{Rstar} \\
    && l_{\star} = \frac{1}{\sigma_{\rm eff}}, \\
    && \sigma_{\rm eff} = \sqrt{\sigma_{\rm W}^{2} + \kappa_{\star}\theta_{\rm Ein}^{2}}.
\end{eqnarray}
The variance of the total magnification $\langle \mu_{\rm tot}^{2} \rangle$ is expressed in the form of 
\begin{eqnarray}
    \langle \mu_{\rm tot}(\boldsymbol{\beta})^{2} \rangle = && \int d^{2}\theta' \int d^{2}\theta'' \frac{1}{(2\pi)^{2} \sqrt{\det(C_{\rm ml}(\boldsymbol{\Tilde{\theta}}) + \sigma_{\rm W}^{2} I)}} \nonumber \\
    && \times \exp \left[-\frac{1}{2} \boldsymbol{u}^{\rm T} (C_{\rm ml}(\boldsymbol{\Tilde{\theta}}) + \sigma_{\rm W}^{2} I)^{-1} \boldsymbol{u} \right], \label{mu_dis_general}
\end{eqnarray}
where $\boldsymbol{\Tilde{\theta}}=\boldsymbol{\theta''} - \boldsymbol{\theta'}$, and the vector $\boldsymbol{u}$ is 
\begin{equation}
    \boldsymbol{u}(\boldsymbol{\theta}', \boldsymbol{\theta}''; \boldsymbol{\beta}) = 
    \begin{pmatrix}
        \theta'_{1} - \beta_{1} - \alpha_{\rm B1}(\boldsymbol{\theta'}) \\
        \theta'_{2} - \beta_{2} - \alpha_{\rm B2}(\boldsymbol{\theta'}) \\
        \theta''_{1} - \beta_{1} - \alpha_{\rm B1}(\boldsymbol{\theta''}) \\
        \theta'_{2} - \beta_{2} - \alpha_{\rm B2}(\boldsymbol{\theta''})
    \end{pmatrix}.
\end{equation}
The matrix $C_{\rm ml}(\boldsymbol{\Tilde{\theta}})$ is given as
\begin{equation}
    C_{\rm ml}(\boldsymbol{\Tilde{\theta}}) = 
    \begin{pmatrix}
        \sigma_{\rm ml}^{2} & 0 & C^{\rm ml}_{13} & C^{\rm ml}_{14} \\
        0 & \sigma_{\rm ml}^{2} &  C^{\rm ml}_{14} & C^{\rm ml}_{24} \\
       C^{\rm ml}_{13} & C^{\rm ml}_{14} & \sigma_{\rm ml}^{2} & 0 \\
       C^{\rm ml}_{14} & C^{\rm ml}_{24} & 0 & \sigma_{\rm ml}^{2}
    \end{pmatrix},
\end{equation}
where each component is
\begin{eqnarray}
    && C^{\rm ml}_{13}(\boldsymbol{\Tilde{\theta}}) = C^{\rm ml}_{\parallel}(\Tilde{\theta}) \cos^{2}\phi + C^{\rm ml}_{\perp} \sin^{2}\phi, \\
    && C^{\rm ml}_{14}(\boldsymbol{\Tilde{\theta}}) = C^{\rm ml}_{\parallel}(\Tilde{\theta}) \cos\phi \sin\phi - C^{\rm ml}_{\perp}(\Tilde{\theta}) \cos\phi \sin\phi ,\\
    && C^{\rm ml}_{24}(\boldsymbol{\Tilde{\theta}}) = C^{\rm ml}_{\parallel}(\Tilde{\theta}) \sin^{2}\phi + C^{\rm ml}_{\perp}(\Tilde{\theta}) \cos^{2}\phi,
\end{eqnarray}
where $\phi$ is the polar angle of $\boldsymbol{\Tilde{\theta}}$ and 
$C^{\rm ml}_{\parallel}(\Tilde{\theta})$ and $C^{\rm ml}_{\perp}(\Tilde{\theta})$ are given by
\begin{eqnarray}
    && C^{\rm ml}_{\parallel}(\Tilde{\theta}) = \kappa_{\star} \theta_{\rm Ein}^{2} \left[\ln \frac{R_{\star}}{\Tilde{\theta}} + \ln \sqrt{1+\frac{\Tilde{\theta}^{2}}{4R_{\star}^{2}}} - \frac{1}{2} \right], \label{C_parallel}\\
    && C^{\rm ml}_{\perp}(\Tilde{\theta}) = \kappa_{\star} \theta_{\rm Ein}^{2} \left[\ln \frac{R_{\star}}{\Tilde{\theta}} + \ln \sqrt{1+\frac{\Tilde{\theta}^{2}}{4R_{\star}^{2}}} + \frac{1}{2} \right]. \label{C_perp}
\end{eqnarray}
The standard deviation is obtained from Eqs.~\eqref{mu_av_general} and \eqref{mu_dis_general}
\begin{equation}
    {\rm Std}[\mu_{\rm tot}(\boldsymbol{\beta})] = \sqrt{\langle \mu_{\rm tot}^{2}(\boldsymbol{\beta})\rangle - \mu_{\rm av}^{2}(\boldsymbol{\beta})}. \label{mu_std_general}
\end{equation}
Note that these analytical estimations are invalid where $\Tilde{\theta} \gg R_{\star}$ and $\Tilde{\theta} \ll \theta_{\rm Ein}^{2}l_{\star}$.
Therefore, this analytic model cannot properly predict the high-magnification tail of the PDF that we study in this paper.

When the background convergence $\kappa_{\rm B}$ and shear $\gamma_{\rm B}$ are constant, these equations are expressed in simpler forms.
The average magnification is 
\begin{equation}
    \mu_{\rm av} = \mu_{\rm B} = \frac{1}{|(1-\kappa_{\rm B})^{2}-\gamma_{\rm B}^{2}|}, \label{mu_av_constant_kg}
\end{equation}
and the variance of the total magnification is
\begin{equation}
    \langle \mu_{\rm tot}^{2} \rangle = \frac{\mu_{\rm av}^{2}}{4\pi} \int d^{2}\Tilde{\Theta} \frac{\exp \left[-\frac{1}{2} \Tilde{\Theta}^{\rm T} \cdot D(\Tilde{\theta}) \cdot \Tilde{\Theta}\right]}{\sqrt{B_{\rm ml}(\Tilde{\theta})}}, \label{mu_dis_constant_kg}
\end{equation}
where $B_{\rm ml}(\Tilde{\theta})$ is defined as
\begin{equation}
    B_{\rm ml}(\Tilde{\theta}) = (\sigma_{\rm ml}^{2} + \sigma_{\rm W}^{2} - C_{\parallel}^{\rm ml}(\Tilde{\theta})) (\sigma_{\rm ml}^{2} + \sigma_{\rm W}^{2} - C_{\perp}^{\rm ml}(\Tilde{\theta})).
\end{equation}
Here $\boldsymbol{\Tilde{\Theta}}$ is defined by $\boldsymbol{\Theta}'' - \boldsymbol{\Theta}'$ where $\boldsymbol{\Tilde{\Theta}}$ is the rescaled coordinate on the lens plane, 
\begin{equation}
    \boldsymbol{\Theta} = 
    \begin{pmatrix}
        |1-\kappa_{\rm B}-\gamma_{\rm B}| \theta_{1}\\
        |1-\kappa_{\rm B}+\gamma_{\rm B}| \theta_{2}
    \end{pmatrix}.
\end{equation}
The components of the two dimensional matrix $D(\Tilde{\theta})$ are
\begin{eqnarray}
    && D_{11} = \frac{1}{4B_{\rm ml}(\Tilde{\theta})} \{2\sigma_{\rm ml}^{2} + 2\sigma_{\rm W}^{2} - (C_{\parallel}^{\rm ml}(\Tilde{\theta}) + C_{\perp}^{\rm ml}(\Tilde{\theta})) \nonumber \\
    && \hspace{30mm} + (C_{\parallel}^{\rm ml}(\Tilde{\theta}) - C_{\perp}^{\rm ml}(\Tilde{\theta})) \cos 2\phi \}, \\
    && D_{12} = D_{21} = \frac{1}{4B_{\rm ml}(\Tilde{\theta})} (C_{\parallel}^{\rm ml}(\Tilde{\theta}) - C_{\perp}^{\rm ml}(\Tilde{\theta})) \sin 2\phi, \\
    && D_{22} = \frac{1}{4B_{\rm ml}(\Tilde{\theta})} \{2\sigma_{\rm ml}^{2} + 2\sigma_{\rm W}^{2} - (C_{\parallel}^{\rm ml}(\Tilde{\theta}) + C_{\perp}^{\rm ml}(\Tilde{\theta})) \nonumber \\
    && \hspace{30mm} - (C_{\parallel}^{\rm ml}(\Tilde{\theta}) - C_{\perp}^{\rm ml}(\Tilde{\theta})) \cos 2\phi \}.
\end{eqnarray}
In this paper, we choose $\phi = 0$ to calculate these values.

Another useful example is when the lens potential is given by Eq.~\eqref{lens_pot_near_cc}.
The deflection angle $\alpha_{\rm B}$ is given by Eq.~\eqref{def_ang_near_cc}.
In this case, the Gaussian integral for $\theta_{2}$ is easy to perform, and we obtain the average magnification by
\begin{eqnarray}
    \mu_{\rm av}(\boldsymbol{\beta}) = && \frac{1}{\sqrt{2\epsilon}(1-\kappa_{0})}\frac{1}{\sqrt{2\pi (\sigma_{\rm ml}^{2} + \sigma_{\rm W}^{2})}} \nonumber \\
    && \times  \int_{0}^{\infty} d\beta'_{1} \frac{1}{\sqrt{\beta'_{1}}} \exp\left[-\frac{(\beta'_{1}-\beta_{1})^{2}}{2(\sigma_{\rm ml}^{2} + \sigma_{\rm W}^{2})}\right]. \label{mu_beta_source_ml_eff}
\end{eqnarray}
Note that Eq.~\eqref{mu_beta_source_ml_eff} is the same as Eq.~\eqref{mu_beta_source_eff} except for the source size.

\begin{figure}
\includegraphics[width=\columnwidth]{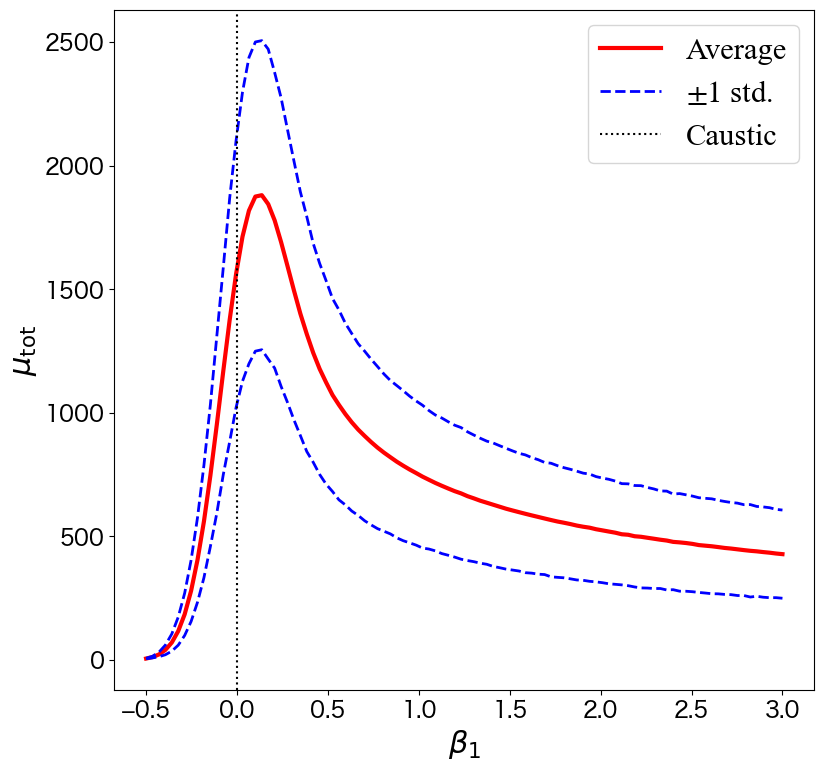}
    \caption{The average (red solid line) and $1\sigma$ standard deviation (blue dashed lines) of the magnifications as a function of the distance from the caustic.
    The vertical dotted line is the position of the caustic. 
    We set $\kappa_{0} = 0.7$, $\epsilon = 10^{-5}$, $\kappa_{\star} = 0.004$, $\theta_{\rm Ein} = 1.0$, $\sigma_{\rm W} = 0.05$, and $R_{\star} = 1500$.
    The average magnification and dispersion are calculated using Eqs.~\eqref{mu_beta_source_ml_eff} and \eqref{mu_dis_constant_kg}.
    Here $\kappa_{\rm B}$ and $\gamma_{B}$ are obtained from Eq.~\eqref{mu_av_constant_kg} assuming $\kappa_{\rm B} = \gamma_{B}$.
    The standard deviation is calculated by Eq.~\eqref{mu_std_general}.
    }
    \label{fig:mu_av_std}
\end{figure}

Figure \ref{fig:mu_av_std} shows an example of the average and 1$\sigma$ standard deviation of the magnification.
We use the lens potential given by Eq.~\eqref{lens_pot_near_cc} with $\kappa_{0} = 0.7$ and $\epsilon = 10^{-5}$.
The average magnification is calculated by Eq.~\eqref{mu_beta_source_ml_eff}, where we set $\kappa_{\star} = 0.004$, $\theta_{\rm Ein} = 1.0$, and $\sigma_{\rm W} = 0.05$.
We follow \citet{2021arXiv210412009D} and set $R_{\star} = 1500$, which is different from Eq.~\eqref{Rstar}.
To calculate the dispersion, we do not directly use Eq.~\eqref{mu_dis_general} due to the computational problem.
Instead, we first compute the convergence and the shear by using the average magnification, Eq.~\eqref{mu_av_constant_kg} assuming $\kappa_{\rm B} = \gamma_{\rm B}$, for each $\beta_{1}$.
Then, we use Eq.~\eqref{mu_dis_constant_kg} to obtain the dispersion of the total magnification.
Even with this simplified assumption, we find that derived values approximately match those obtained from the full calculation.
Finally, we obtain the standard deviation from Eq.~\eqref{mu_std_general}.
To avoid the violation of the regimes of $\Tilde{\theta} \gg R_{\star}$ and $\Tilde{\theta} \ll \theta_{\rm Ein}^{2}l_{\star}$, we use the following equation instead of Eqs.~\eqref{C_parallel} and \eqref{C_perp} as introduced by \citet{2021arXiv210412009D},
\begin{eqnarray}
    C^{\rm ml}_{\rm X,reg}(\Tilde{\theta}) && = \sigma_{\rm ml}^{2}\ {\rm sgn} (C^{\rm ml}_{\rm X}(\Tilde{\theta})) \nonumber \\
    && \times \left[ 1+\left( \frac{\sigma_{\rm ml}^{2}}{|C^{\rm ml}_{\rm X}(\Tilde{\theta})| e^{-(\Tilde{\theta}/(\nu R_{\star}))^{2}}}\right)^{n} \right]^{-1/n},
\end{eqnarray}
where we apply $n =10$ and $\nu = 1$.
We use the Monte Carlo algorithm {\sc vegas} \citep{1978JCoPh..27..192L, 2021JCoPh.43910386L} to compute the integration.

\section{High-magnification tail of probability distribution function} \label{sec:analytical_pdf}
In this section, we describe how we model the high magnification tail of the PDF.

We consider the case where microlenses exist in a constant smooth background.
The total average convergence (mass) and the shear are denoted by $\kappa_{\rm tot}$ and $\gamma_{\rm tot}$, respectively.
We assume the relation $\kappa_{\rm tot} = \gamma_{\rm tot}$ since the ultra-high magnification events are observed near the critical curves of galaxy clusters, whose mass distribution can often be approximated by the spherical isothermal density distribution.
The average convergence of microlenses is denoted by $\kappa_{\star}$, and the mass fraction of the microlenses $f_{\star}$ is defined by $\kappa_{\star} = f_{\star} \kappa_{\rm tot}$.
Here we assume that individual masses of microlenses are the same and they are uniformly distributed randomly on the lens plane.
Note that these assumptions might not be accurate since the distribution of microlenses should follow a mass function and some of them might form binary systems, which we discuss in Sec.~\ref{sec:discussion_conclusion}.
The original Einstein radii of all microlenses are the same, which is denoted by $\theta_{\rm Ein}$, and the number of microlenses, $N_{\star}^{\rm tot}$, is proportional to the mass fraction of the microlenses $f_{\star}$.
The average magnification does not depend on the mass fraction $f_{\star}$ but on the total mass, as is analytically shown by \citet{2021arXiv210412009D} (Eq.~\eqref{mu_av_constant_kg}), and is expressed by 
\begin{equation}
    \mu_{\rm av} = \frac{1}{|(1-\kappa_{\rm tot})^{2}-\gamma_{\rm tot}^{2}|} = \frac{1}{|1-2\kappa_{\rm tot}|}.
\end{equation}
To be precise, while the background convergence would be $\kappa_{\rm B} = \kappa_{\rm tot} - \kappa_{\star}$ as we fix the total convergence, the simulation conducted by {\sc Gerlumph} \citep{2014ApJS..211...16V} also supports the above relation. 
Note that the average magnification can be decomposed into the tangential magnification $\mu_{\rm t, av}$ and radial magnification $\mu_{\rm r, av}$, 
where they satisfy $\mu_{\rm t, av} = \mu_{\rm av}$ and $\mu_{\rm r, av} = 1$ since we adopt $\kappa_{\rm tot} = \gamma_{\rm tot}$.

We study the parameter dependence of the high magnification tail of the PDF on the average magnification $\mu_{\rm av}$, which depends solely on the total convergence, and the mass fraction of microlenses $f_{\star}$.
For convenience, we introduce the normalized magnification 
\begin{equation}
    r = \frac{\mu_{\rm tot}}{\mu_{\rm av}}
\end{equation}
and focus on the high magnification regime, $r \gtrsim 10$.
Note that, since the average magnification is determined by the distance from the macro-critical curve or caustic, {\it e.g.}, Eq.~\eqref{mu_av_general}, our model allows us to study how the high-magnification events appear near the macro-critical curve, as discussed in Sec.~\ref{sec:application}.

We find that the high magnification tail of the PDF can well be modeled as 
\begin{equation}
    \frac{dP}{d\log_{10}{r}} \propto N_{\star}^{\rm indep} \sqrt{\mu_{\rm av}}\  r^{-2} S(r; r_{\rm max}), \label{pdf_model_eq}
\end{equation}
where the $N_{\star}^{\rm indep}$ denotes the independent number of micro-critical curves, and the $S(r; r_{\rm max})$ is the suppression function of the PDF above the maximum magnification $r_{\rm max}$ due to the finite source size effect.
In the case of point sources, $S(r; r_{\rm max})=1$.
The factor $\sqrt{\mu_{\rm av}}$ expresses the expansion of the length of each critical curve (and caustic) under the smooth background as shown in Sec.~\ref{subsec:microlens_in_smooth}.
The $r^{-2}$ dependence appears from the width around the critical curve where the magnification takes between $\log_{10}r$ and $\log_{10}r+d\log_{10}r$, as shown in Eq.~\eqref{pdf_tail_mu2}.
To summarize, the first three components on the right-hand side, $N_{\star}^{\rm indep} \sqrt{\mu_{\rm av}}\  r^{-2}$, calculate the total area where the magnification is between $\log_{10}r$ and $\log_{10}r+d\log_{10}r$, and the finite source size effect is incorporated in $S(r; r_{\rm max})$.

The independent micro-critical curve is a micro-critical curve created by a microlens that has the nearest neighbor distance to other microlenses larger than the size of its micro-critical curve.
To quantitatively obtain the independent number of the micro-critical curves, we first derive the distribution of nearest inter-microlens distances $\theta_{\rm nid}$ of the randomly distributed microlenses on the lens plane.

Considering the case where microlenses are distributed randomly on a two-dimensional lens plane with the number density $n_{\star}$, which is given by 
\begin{equation}
    n_{\star} = \frac{f_{\star}\kappa_{\rm tot}}{\pi \theta_{\rm Ein}^{2}}.\label{n_star}
\end{equation}
The mean number of the microlenses in an area $S$ is given by $\bar{N}_{\star} = n_{\star} S$.
Assuming that the number of microlenses follows the Poisson distribution, we can express the probability of the number of microlenses inside the area $S$ as
\begin{equation}
    P(N_{\star}; S) = \frac{1}{N_{\star}!} e^{-n_{\star}S} (n_{\star} S)^{N_{\star}}.
\end{equation}
To obtain the distribution of the nearest neighbor distance from a single microlens, we consider the distribution satisfying the following two conditions.
The first condition is that no microlenses exist within a circle of the radius $\theta_{\rm nid}$ from a microlens.
The second condition is that at least one microlens exists in a small area of radius $\theta_{\rm nid}$ and $\theta_{\rm nid}+d\theta_{\rm nid}$.
The probability satisfying the first condition is 
\begin{equation}
    P(N_{\star}=0; S = \pi \theta_{\rm nid}^{2}) = e^{-n_{\star}\pi \theta_{\rm nid}^{2}}.
\end{equation}
The probability satisfying the second condition is 
\begin{eqnarray}
    1 - P(N_{\star} = 0; S = 2\pi \theta_{\rm nid} d\theta_{\rm nid}) &=& 1-e^{-2n_{\star}\pi \theta_{\rm nid} d\theta_{\rm nid}} \nonumber\\
    &\simeq& 2 n_{\star} \pi \theta_{\rm nid} d\theta_{\rm nid}.
\end{eqnarray}
Therefore, the probability of satisfying both the conditions is 
\begin{equation}
    \frac{dQ}{d\theta_{\rm nid}} = 2n_{\star}\pi \theta_{\rm nid}  e^{-n_{\star}\pi \theta_{\rm nid}^{2}},
\end{equation}
from which we find that the nearest inter-microlens distance follows the Rayleigh distribution.
The mean of the nearest inter-microlens distance is
\begin{equation}
    \bar{\theta}_{\rm nid} = \int_{0}^{\infty} \theta_{\rm nid} \frac{dQ}{d\theta_{\rm nid}} d\theta_{\rm nid} = \frac{1}{2\sqrt{n_{\star}}}. \label{mean_theta_id}
\end{equation}

Now we can calculate the independent number of the micro-critical curves, $N_{\star}^{\rm indep}$.
Since the typical size of each micro-critical curve is $\sqrt{\mu_{\rm av}} \theta_{\rm Ein}$, the probability of a micro-critical curve to be independent can be calculated as
\begin{eqnarray}
    Q(\theta_{\rm nid} > \sqrt{\mu_{\rm av}}\theta_{\rm Ein}) &=& \int_{\sqrt{\mu_{\rm av}}\theta_{\rm Ein}}^{\infty} \frac{dQ}{d\theta_{\rm nid}} d\theta_{\rm nid} \nonumber \\
    &=& \exp(-f_{\star} \kappa_{\rm tot} \mu_{\rm av}).
\end{eqnarray}
Therefore, the number of the independent micro-critical curve is 
\begin{eqnarray}
    N_{\star}^{\rm indep} &=& N_{\star}^{\rm tot} Q(\theta_{\rm nid} > \sqrt{\mu_{\rm av}}\theta_{\rm E}) \nonumber\\
    &\propto& f_{\star} \kappa_{\rm tot} \exp(-f_{\star}\kappa_{\rm tot} \mu_{\rm av}). \label{Nstar_ind} 
\end{eqnarray}
When the fraction of the microlenses is sufficiently small, the nearest neighbor distance is almost always larger than the size of each micro-critical curve, which results in $Q(\theta_{\rm nid} > \sqrt{\mu_{\rm av}} \theta_{\rm Ein}) \simeq 1$ and $N_{\star}^{\rm indep} \simeq N_{\star}^{\rm tot}$.
As the fraction of the microlenses or the average magnification becomes larger, the typical nearest inter-microlens distance becomes smaller than the typical size of each micro-critical curve, and therefore the number of the independent micro-critical curves becomes smaller than the total number of the microlenses, $N_{\star}^{\rm indep} < N_{\star}^{\rm tot}$.
We call the former regime the ``linear regime'', and the latter regime the ``nonlinear regime''.
The boundary of the linear and the nonlinear regimes can be defined by comparing the mean of the nearest inter-microlens distance $\bar{\theta}_{\rm nid}$ given by Eq.~\eqref{mean_theta_id} and the typical size of each micro-critical curve $\sqrt{\mu_{\rm av}} \theta_{\rm Ein}$, resulting 
\begin{equation}
    f_{\star} \kappa_{\rm tot} \mu_{\rm av} \simeq 1. \label{lin_nonlin_condition}
\end{equation}
In the linear regime where $f_{\star} \kappa_{\rm tot} \mu_{\rm av} \lesssim 1$, the number of the independent micro-critical curves is proportional to the fraction of the microlenses, $N_{\star}^{\rm indep} \propto f_{\star}$, as can be understood from Eq.~\eqref{Nstar_ind}.
On the other hand, the number of independent micro-critical curves is exponentially suppressed in the nonlinear regime.
Considering the limit of $f_{\star} \rightarrow 1$, all matter components are included in the microlenses, which in turn approaches the smooth matter distribution.
This is why the suppression occurs in the nonlinear regime.

\begin{figure}
\includegraphics[width=\columnwidth]{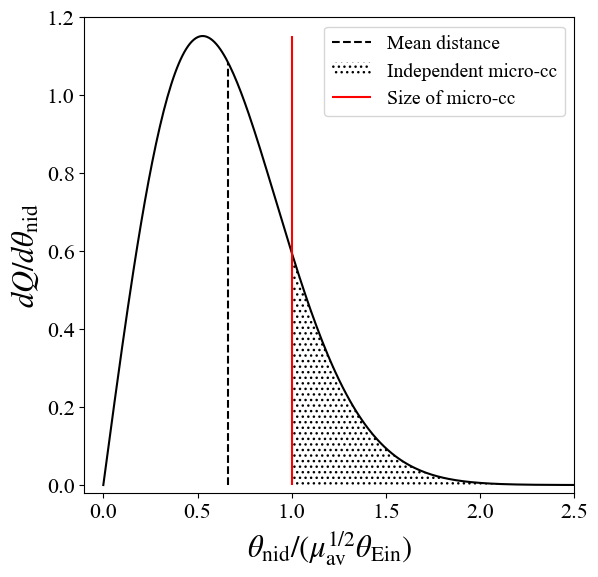}
    \caption{Distribution of the nearest inter-microlens distance (black solid line) and the size of the micro-critical curve (vertical red solid line).
    The mean of the nearest inter-microlens distance is plotted by the vertical black dashed line.
    The dotted region is where the microlenses are independent.
    Here we set $\theta_{\rm Ein} = 0.2$, $\kappa_{\rm tot} = 0.48$, $f_{\star} = 0.15$.
    }
    \label{fig:rayleigh_theta_id}
\end{figure}

In Fig.~\ref{fig:rayleigh_theta_id}, we plot the distribution of the nearest inter-microlens distance (black solid line) and compare it with the size of the micro-critical curve (vertical red solid line). 
The microlenses whose distance to the nearest neighbor is larger than the size of the micro-critical curves (black dotted region) produce the independent micro-critical curves.
The vertical black dashed line is the mean of the nearest inter-microlens distance.
In this case, only a small fraction of the microlenses are independent since the mean of the nearest inter-microlens distance is smaller than the size of the micro-critical curve, {\it i.e.}, the nonlinear regime. 

In the case of a point source where the suppression factor is $S(r;r_{\rm max}) = 1$, we can model the high magnification tail of the PDF by combining all these results to find
\begin{equation}
    \frac{dP}{d\log_{10} r} \propto f_{\star} \kappa_{\rm tot} \exp(-f_{\star} \kappa_{\rm tot} \mu_{\rm av}) \sqrt{\mu_{\rm av}} r^{-2}.
    \label{pdf_model_eq_s1_mod}
\end{equation}
By integrating PDF above a threshold magnification $r_{\rm th}$, we can obtain
\begin{eqnarray}
    && P^{\rm PS}(r>r_{\rm th}) = \int_{r_{\rm th}}^{\infty} \frac{dP}{d\log_{10} r} d\log_{10} r \nonumber\\ 
    && \hspace{12mm} \propto f_{\star} \kappa_{\rm tot} \sqrt{\mu_{\rm av}} \exp(-f_{\star} \kappa_{\rm tot} \mu_{\rm av}) \left(\frac{1}{r_{\rm th}}\right)^{2}, \label{P_ps_XY}
\end{eqnarray}
where PS denotes the point source.
It is found that, by multiplying $\sqrt{\mu_{\rm av}}$ in both sides, Eq.~\eqref{P_ps_XY} can be expressed as
\begin{equation}
    Y \propto X\exp(-X) \label{eq:XY},
\end{equation}
where $X = f_{\star} \kappa_{\rm tot} \mu_{\rm av}$ and $Y =  P^{\rm PS}(r>r_{\rm th}) \sqrt{\mu_{\rm av}}$.
The boundary of the linear and nonlinear regimes is $X \simeq 1$.
We find for the first time that, by considering this parameter combination, the dependence of the high magnification PDF on $f_{\star}$ and $\mu_{\rm av}$ can be examined in a unified manner covering both the linear and nonlinear regimes.
We leave the discussion on the suppression factor due to the finite source size in Sec.~\ref{sec:comparison_with_simulation}.

\section{Comparison with simulation} \label{sec:comparison_with_simulation}
To test our model shown in Sec.~\ref{sec:analytical_pdf}, we conduct detailed ray-tracing simulations using our new code {\sc CCtrain},  which adopts the same algorithm for solving the lens equation as that used in {\sc Glafic} \citep{2010PASJ...62.1017O}. We note that {\sc CCtrain} has also been used in \citet{2024arXiv240408094W} to study the magnification pattern near macro-critical curves. In addition to the adaptive grid to efficiently solve the lens equation, {\sc CCtrain} uses a hierarchical tree algorithm to speed up the calculation of deflection angles from a large number of microlenses \cite{1999JCoAM.109..353W}. Since magnifications are derived from the second derivatives of the lens potential at each source position as $\mu = 1/\{(1-\kappa)^{2}-\gamma^{2}\}$, where $\kappa$ and $\gamma$ are the convergence and shear at the image position, respectively, magnifications derived from {\sc CCtrain} correspond to those for point sources without any finite source size effects.

In addition to our detailed ray-tracing simulations with {\sc CCtrain}, we adopt public magnification maps provided by {\sc Gerlumph} \citep{2014ApJS..211...16V} for our study. The {\sc Gerlumph} simulations are based on the inverse ray-tracing simulation, for which the magnification is calculated by the number of the inverse rays entered in a single pixel on the source plane.
Therefore, the grid size of the source plane can be interpreted as an effective source size for the {\sc Gerlumph} simulations. 

For both {\sc CCtrain} and {\sc Gerlumph}, we consider the situation where the microlenses are distributed randomly in the background with constant convergence and shear which satisfies $\kappa_{\rm tot} = \gamma_{\rm tot}$, which is the same as in Sec.~\ref{sec:analytical_pdf}.
Note that {\sc Gerlumph} uses the smooth fraction $s$, which is related to the fraction of the microlenses by $f_{\star} = 1-s$. 
For {\sc CCtrain}, for a given parameter setup, we change the distribution of the microlenses and obtain many realizations as summarized in Tab.~\ref{tab:glafic_realization}. 
In the case of {\sc Gerlumph}, we adopt the public data provided on the website,  https://gerlumph.swin.edu.au. 
The realizations we use are listed in Tab.~\ref{tab:gerlumph_realization}.

Using these realizations, we can obtain the PDF of the magnification.
Note again that we study the distribution of the total magnification, the sum of the multiple images.
We focus on the high-magnification region $\mu_{\rm tot} \gtrsim 10 \mu_{\rm av}$, which corresponds to $r \gtrsim r_{\rm th}$ with the threshold of the normalized magnification $r_{\rm th} \simeq 10$.

\begin{table*}[ht]
    \centering
    \caption{The simulation setup conducted by {\sc CCtrain}. $l_{x}$ and $l_{y}$ are the lengths of the box size in the $x$- and $y$-directions, respectively. The box size and the resolution on the lens plane are in the unit of the Einstein radius of a point mass lens.}
    \begin{tabular}{c|c|c|c|c|c}
        $\kappa_{\rm tot}$ & $\mu_{\rm av}$ & $f_{\star}$ & $(l_{x}, l_{y})$ & resolution (lens plane) &  realization \\
        \hline \hline
        0.45 & 10 & 0.0015625 & (15, 2)    & 0.003125 & 100,000 \\
        0.45 & 10 & 0.003125  & (15, 2)    & 0.003125 & 100,000 \\
        0.45 & 10 & 0.00625   & (15, 2)    & 0.003125 & 100,000 \\
        0.45 & 10 & 0.0125    & (15, 2)    & 0.003125 & 100,000 \\
        0.45 & 10 & 0.025     & (20, 3)    & 0.003125 & 100,000 \\
        0.45 & 10 & 0.05      & (20, 3)    & 0.003125 & 100,000 \\
        0.45 & 10 & 0.10      & (30, 3)    & 0.003125 & 61,549 \\
        0.45 & 10 & 0.25      & (50, 4)    & 0.003125 & 349 \\
        0.49 & 50 & 0.0003125 & (9, 2)     & 0.0125   & 100,000 \\
        0.49 & 50 & 0.000625  & (12, 2)    & 0.0125   & 100,000 \\
        0.49 & 50 & 0.00125   & (18, 2)    & 0.0125   & 100,000 \\
        0.49 & 50 & 0.0025    & (25, 2)    & 0.0125   & 100,000 \\
        0.49 & 50 & 0.005     & (35, 2)    & 0.0125   & 100,000 \\
        0.49 & 50 & 0.01      & (50, 2)    & 0.0125   & 100,000 \\
        0.49 & 50 & 0.02      & (70, 2)    & 0.0125   & 100,000 \\
        0.49 & 50 & 0.04      & (100, 2)   & 0.0125   & 100,000 \\
        0.49 & 50 & 0.056     & (119, 4)   & 0.0125   & 10,000 \\
        0.49 & 50 & 0.08      & (140, 4)   & 0.0125   & 10,000 \\
        0.49 & 50 & 0.112     & (166, 4)   & 0.0125   & 10,000 \\
        0.49 & 50 & 0.16      & (200, 4)   & 0.0125   & 10,000 \\
        0.49 & 50 & 0.32      & (280, 5.6) & 0.0125   & 3,926 \\
    \end{tabular}
    \label{tab:glafic_realization}
\end{table*}

\begin{table*}[ht]
    \centering
    \caption{The realizations of {\sc Gerlumph} we use in this study. The resolution on the source plane is in the unit of the Einstein radius of a point mass lens.}
    \begin{tabular}{c|c|c|c|c}
        $\kappa_{\rm tot}$ & $\mu_{\rm av}$ & $f_{\star}$ & resolution (source plane) & realization \\
        \hline \hline
        0.30 & 2.50 & 0.01, 0.05, 0.1, 0.2, 0.3, 0.4, 0.5, 0.6, 0.7, 0.8, 0.9, 1 & 0.0025 & 100,000,000 (each)\\
        0.33 & 2.94 & 0.01, 0.05, 0.1, 0.2, 0.3, 0.4, 0.5, 0.6, 0.7, 0.8, 0.9, 1 & 0.0025 & 100,000,000 (each)\\
        0.36 & 3.57 & 0.01, 0.05, 0.1, 0.2, 0.3, 0.4, 0.5, 0.6, 0.7, 0.8, 0.9, 1 & 0.0025 & 100,000,000 (each)\\
        0.37 & 3.85 & 0.01, 0.05, 0.1, 0.2, 0.3, 0.4, 0.5, 0.6, 0.7, 0.8, 0.9, 1 & 0.0025 & 100,000,000 (each)\\
        0.38 & 4.12 & 0.01, 0.05, 0.1, 0.2, 0.3, 0.4, 0.5, 0.6, 0.7, 0.8, 0.9, 1 & 0.0025 & 100,000,000 (each)\\
        0.39 & 4.55 & 0.01, 0.05, 0.1, 0.2, 0.3, 0.4, 0.5, 0.6, 0.7, 0.8, 0.9, 1 & 0.0025 & 100,000,000 (each)\\
        0.40 & 5.00 & 0.01, 0.05, 0.1, 0.2, 0.3, 0.4, 0.5, 0.6, 0.7, 0.8, 0.9, 1 & 0.0025 & 100,000,000 (each)\\
        0.41 & 5.56 & 0.01, 0.05, 0.1, 0.2, 0.3, 0.4, 0.5, 0.6, 0.7, 0.8, 0.9, 1 & 0.0025 & 100,000,000 (each)\\
        0.42 & 6.25 & 0.01, 0.05, 0.1, 0.2, 0.3, 0.4, 0.5, 0.6, 0.7, 0.8, 0.9, 1 & 0.0025 & 100,000,000 (each)\\
        0.43 & 7.14 & 0.01, 0.05, 0.1, 0.2, 0.3, 0.4, 0.5, 0.6, 0.7, 0.8, 0.9, 1 & 0.0025 & 100,000,000 (each)\\
        0.44 & 8.33 & 0.01, 0.05, 0.1, 0.2, 0.3, 0.4, 0.5, 0.6, 0.7, 0.8, 0.9, 1 & 0.0025 & 100,000,000 (each)\\
        0.45 & 10.0 & 0.01, 0.05, 0.1, 0.2, 0.3, 0.4, 0.5, 0.6, 0.7, 0.8, 0.9, 1 & 0.0025 & 100,000,000 (each)\\
        0.46 & 1.25 & 0.01, 0.05, 0.1, 0.2, 0.3, 0.4, 0.5, 0.6, 0.7, 0.8, 0.9, 1 & 0.0025 & 100,000,000 (each)\\
        0.47 & 16.7 & 0.01, 0.05, 0.1, 0.2, 0.3, 0.4, 0.5, 0.6, 0.7, 0.8, 0.9, 1 & 0.0025 & 100,000,000 (each)\\
        0.48 & 25.0 & 0.01, 0.05, 0.1, 0.2, 0.3, 0.4, 0.5, 0.6, 0.7, 0.8, 0.9, 1 & 0.0025 & 100,000,000 (each)\\
    \end{tabular}
    \label{tab:gerlumph_realization}
\end{table*}

In what follows, we consider two cases for testing our model, cases of a point source and a finite source size, which are tested by {\sc CCtrain} and {\sc Gerlumph}, respectively. 
In Sec.~\ref{subsec:point_source}, we describe the point source case, in which we assume that the suppression factor $S(r;r_{\rm max})$ can be ignored.
We then study the case with a finite source size in Sec.~\ref{subsec:finite_source}.

\subsection{Point source} \label{subsec:point_source}
In the case of a point source, we can ignore the suppression term and set $S(r; r_{\rm max}) = 1$ in Eq.~\eqref{pdf_model_eq}.
As shown in Eq.~\eqref{eq:XY}, the dependence of the parameters on the high magnification tail of the PDF can be absorbed with $X$ and $Y$. Thus all data should ideally be located along a single line in the $X$-$Y$ plane.
We introduce two fitting parameters, $A_{0}$ and $B_{0}$, where $A_{0}$ is introduced to determine the normalization and $B_{0}$ is introduced to take account of the uncertainty in Eq.~\eqref{lin_nonlin_condition}.
Therefore, $B_{0}$ should be $\mathcal{O}(1)$.
With $A_{0}$ and $B_{0}$, our model of the magnification PDF is expressed in the form of 
\begin{eqnarray}
    &&P^{\rm PS}(r>r_{\rm th}) \sqrt{\mu_{\rm av}} \nonumber \\
    && \hspace{5mm} = \frac{A_{0}}{2.4} f_{\star} \kappa_{\rm tot} \mu_{\rm av} \exp(-B_{0} f_{\star} \kappa_{\rm tot} \mu_{\rm av}) \left(\frac{10}{r_{\rm th}}\right)^{2}. \label{P_ps_A0_B0}
\end{eqnarray}
We determine these parameters by fitting the simulation data obtained from {\sc CCtrain} with the threshold magnification $r_{\rm th} = 10$.

\begin{figure}
\includegraphics[width=\columnwidth]{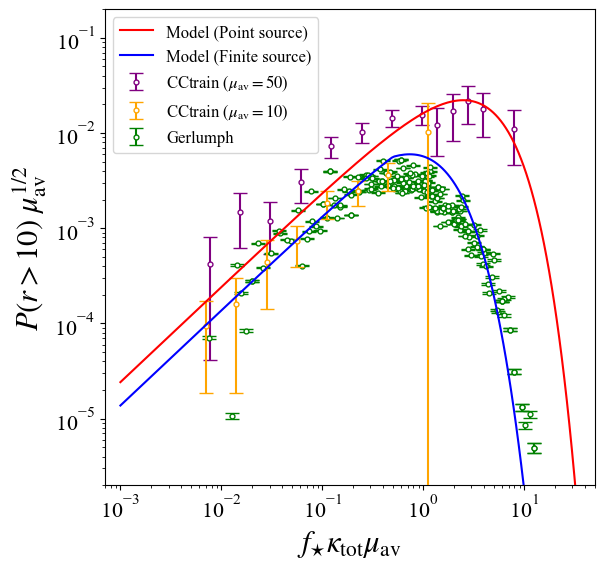}
    \caption{Comparing the integrated PDF between the simulation and our model.
    The purple and orange dots are obtained by the {\sc CCtrain} with $\kappa_{\rm tot} = 0.49$ and $0.45$, respectively.
    The green dots are the simulation result of the {\sc Gerlumph}.
    The realizations are listed in Tab.~\ref{tab:glafic_realization} and \ref{tab:gerlumph_realization}. 
    The red line is the fitting result of our model in the case of the point source with fitting parameters $A_{0} = 0.058$ and $B_{0} = 0.402$.
    The blue line is the fitting result of our model with the finite source case with fitting parameter $C_{0} = 2.0$.
    }
    \label{fig:model_simdata}
\end{figure}

In Fig.~\ref{fig:model_simdata}, we show the result of the fitting. 
We find that the best fitting parameters are $A_{0} = 0.058$ and $B_{0} = 0.402$.
It is seen that the PDF shows the turnover between linear and nonlinear regimes, which is well reproduced by our model.
As expected, the simulation data with different parameters are approximately along a single line in this plane, and the best-fitting value of $B_{0}$ is $\mathcal{O}(1)$.

\subsection{Finite source size} \label{subsec:finite_source}
Taking the finite source size into account, our model of the high-magnification tail of the PDF is given by
\begin{equation}
    \frac{dP}{d\log_{10}{r}} = 2\ln 10\ P^{\rm PS}(r>r_{\rm th})S(r;r_{\rm max})\left(\frac{r}{r_{\rm th}}\right)^{-2}. \label{dp_dlog10r_S}
\end{equation}
Here we discuss the suppression function $S(r; r_{\rm max})$ due to the finite source size, using the simulation data obtained from {\sc Gerlumph}.

There is a maximum magnification due to the averaging effect in the case of the finite source size, as described in Sec.~\ref{subsec:mu_near_cc}.
The resolution on the source plane in {\sc Gerlumph}, can be interpreted as the effective source size, {\it{i.e.}}, $\sigma_{\rm W} = 0.0025 \theta_{\rm Ein}$.
In the linear regime, where the micro-critical curves and micro-caustics are independent of each other, the maximum magnification can be estimated from Eq.~\eqref{mag_near_caustic_w_bg} as 
\begin{equation}
    r_{\rm max} \simeq \sqrt{\frac{\theta_{\rm Ein}}{\sqrt{\mu_{\rm av}}\sigma_{\rm W}}}, \label{rmax_linear}
\end{equation}
where we ignore the $\mathcal{O}(1)$ prefactor.
In the nonlinear regime, following \citet{2017ApJ...850...49V}, the size of the micro-critical curves can be estimated by the average of the nearest inter-microlens distance.
From Eqs.~\eqref{n_star} and \eqref{mean_theta_id}, the distance is expressed as 
\begin{equation}
    \bar{\theta}_{\rm nid} \simeq \frac{\theta_{\rm Ein}}{\sqrt{f_{\star}\kappa_{\rm tot}}},
\end{equation}
where we ignore the factor of two.
Therefore, the maximum magnification in the nonlinear regime can be obtained by replacing $\sqrt{\mu_{\rm av}}$ to $1/\sqrt{f_{\star}\kappa_{\rm tot}}$ in Eq.~\eqref{rmax_linear}.
We obtain the maximum magnification in the nonlinear regime
\begin{equation}
    r_{\rm max} \simeq \sqrt{\frac{\theta_{\rm Ein}}{\sqrt{\mu_{\rm av}}\sigma_{\rm W}}} \left(\frac{1}{f_{\star}\kappa_{\rm tot} \mu_{\rm av}}\right)^{\frac{3}{4}}. \label{rmax_nonlinear}
\end{equation}
Combining these results, we use the following expression of the maximum magnification
\begin{equation}
    r_{\rm max} \simeq \sqrt{\frac{\theta_{\rm Ein}}{\sqrt{\mu_{\rm av}} \sigma_{\rm W}}} \min\left(1,\ (C_{0} f_{\star}\mu_{\rm av}\kappa_{\rm tot})^{-\frac{3}{4}} \right), \label{rmax_lin_nonlinear}
\end{equation}
where we introduce an $\mathcal{O}(1)$ fitting parameter $C_{0}$ to take account of the uncertainty of the estimation in Eq.~\eqref{rmax_nonlinear}.

\begin{figure}
\includegraphics[width=\columnwidth]{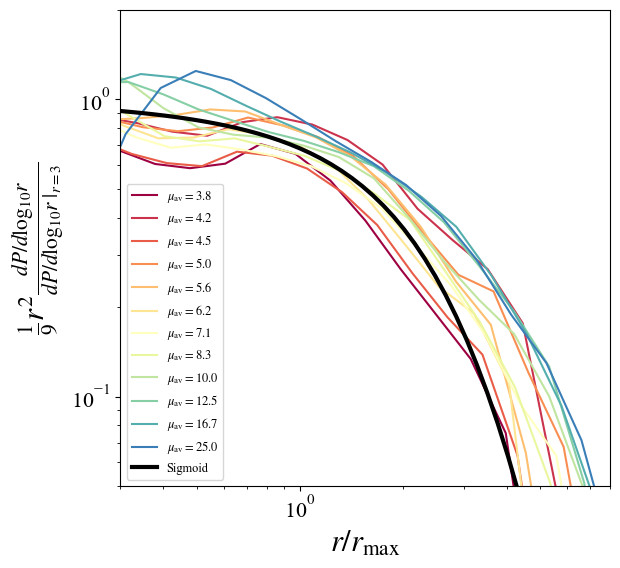}
    \caption{The suppression of the PDF obtained from {\sc Gerlumph} data and the sigmoid function.
    The color lines correspond to the different average magnification (total convergence). 
    Here we use the data with $f_{\star} = 0.2$.
    The black line shows our sigmoid model of the suppression with the fitting parameter $C_{0}=2.0$.
    }
    \label{fig:sigmoid}
\end{figure}

By examining the {\sc Gerlumph} results, we find that the sigmoid function is suitable for describing the suppression,
\begin{equation}
    S(r; r_{\rm max}) = \frac{1+e^{-1}}{1+\exp\left(\frac{r-r_{\rm max}}{r_{\rm max}}\right)}.
\end{equation}
The above function becomes one when the maximum magnification is infinity, {\it i.e.,} in the point source case.
In Fig.~\ref{fig:sigmoid}, we show the result of the fitting.
Here we use the relation
\begin{equation}
    S(r;r_{\rm max}) = r^{2}\frac{dP}{d\log_{10} r} \left(\left.r_{0}^{2}\frac{dP}{d\log_{10} r}\right|_{r=r_{0}} \right)^{-1}, \label{S_obtain}
\end{equation}
which can be obtained from Eq.~\eqref{dp_dlog10r_S}, and set the $x$ axis as the magnification normalized by the maximum magnification using Eq.~\eqref{rmax_lin_nonlinear} and the $y$ axis as the right-hand side of the Eq.~\eqref{S_obtain} with $r_{0} = 3$.
Fig.~\ref{fig:sigmoid} indicates that the sigmoid function with the maximum magnification defined by Eq.~\eqref{rmax_lin_nonlinear} can well describe the simulation results with the fitting parameter $C_{0}=2.0$. 
In Fig.~\ref{fig:model_simdata}, we can check the consistency of the PDF between our model including the suppression (blue line) and the simulation data (green points).
Strictly speaking, the data points in the case of the finite source size do not align in a single line as is different from the point source case, since the additional $\mu_{\rm av}$ dependence is incorporated in the maximum magnification that is used for the suppression function.
However, we find that this effect is rather small and the data points are aligned almost in a single line.
We note that we set the typical average magnification $\mu_{\rm av} = 6.25\ (\kappa_{\rm tot} = 0.42)$ and obtain the maximum magnification to plot our theoretical prediction in Fig.~\ref{fig:model_simdata}.

Combining all the results shown here, we find that the high magnification tail of the PDF can be expressed as
\begin{eqnarray}
    &&\frac{dP}{d\log_{10} r} = 2\ln 10\ P^{\rm PS}(r>r_{\rm th}) \nonumber \\
    && \hspace{20mm} \times \frac{1+e^{-1}}{1+\exp\left(\frac{r-r_{\rm max}}{r_{\rm max}}\right)}\left(\frac{r}{r_{\rm th}}\right)^{-2}. \label{eq:model_summary_1}
\end{eqnarray}
The fitting parameters are $A_{0} = 0.058$, $B_{0} = 0.402$, and $C_{0} = 2.0$.
We find that this model well describes the simulation results. 

By integrating the PDF above in the case of the finite source size, we can obtain the following ratio
\begin{eqnarray}
    &&\frac{P^{\rm FS}(r>r_{\rm th})}{P^{\rm PS}(r>r_{\rm th})} \nonumber \\
    && \hspace{2mm} = 2\ln 10\ r_{\rm th}^{2} \int_{r_{\rm th}}^{\infty} \frac{1}{r^{2}} \frac{1+e^{-1}}{1+\exp\left(\frac{r-r_{\rm max}}{r_{\rm max}}\right)} d\log_{10} r \nonumber \\
    && \hspace{2mm} \simeq \left(1-\frac{r_{\rm th}}{r_{\rm max}}\right) e^{-\frac{r_{\rm th}}{r_{\rm max}}} \nonumber\\
    && \hspace{7mm} + \left(\frac{r_{\rm th}}{r_{\rm max}}\right)^{2} \left\{-{\rm CI}\left(\frac{r_{\rm th}}{r_{\rm max}}\right) + {\rm SI}\left(\frac{r_{\rm th}}{r_{\rm max}}\right)\right\},
\end{eqnarray}
where CI and SI are the hyperbolic cosine integral and the hyperbolic sine integral, respectively.
In the limit of the point source case, {\it i.e.}, $r_{\rm max} \gg r_{\rm th}$, the first term in the right-hand side dominates and reduces to one as expected.

\section{Application to Icarus-like system} \label{sec:application}
In this section, we use our model to predict the event rate of the high magnification events in an Icarus-like system and the probability distribution of their observed positions.
In Sec.~\ref{subsec:Icarus_system}, we show the basic properties of the Icarus system studied in \citet{2018PhRvD..97b3518O}.
In Sec.~\ref{subsec:total_pdf}, we show how we model the total PDF that includes the middle region {\it{i,e.}}, around the average magnification, in addition to the high-magnification tail.
Using our total PDF, we show the prediction of the high-magnification events in the Icarus-like system in Sec.~\ref{subsec:prediction}.

\subsection{Icarus system} \label{subsec:Icarus_system}
As described in Sec.~\ref{sec:intro}, Icarus is an individual star at redshift $z=1.49$ highly magnified by MACS J1149 galaxy cluster at redshift $z=0.544$.
It is estimated to be magnified by a factor of more than two thousands.

From the analytic formula Eq.~\eqref{mu_beta_source_ml_eff} and the previous analysis in \citet{2018PhRvD..97b3518O}, the average magnification $\mu_{\rm av}$ as a function of the distance from the macro-caustic of the galaxy cluster can be expressed as
\begin{eqnarray}
    \mu_{\rm av}(\beta) =&& \mu_{\rm h} \mu_{\rm r} \frac{1}{\sqrt{2\pi(\sigma_{\rm ml}^{2} + \sigma_{\rm W}^{2})}} \nonumber \\
    &&\times \int_{0}^{\infty} d\beta' \sqrt{\frac{\beta_{0}}{\beta'}} \exp\left[-\frac{(\beta'-\beta)^{2}}{2(\sigma_{\rm ml}^{2} + \sigma_{\rm W}^{2})}\right], \label{Icarus_muav}
\end{eqnarray}
with $\mu_{\rm h} = 13$, $\mu_{\rm r} = 3$, and $\beta_{0}=0.045\ {\rm arcsec}$. 
The tangential magnification can be expressed by
\begin{equation}
    \mu_{\rm t}(\theta) = \mu_{\rm h} \sqrt{\frac{\beta_{0}}{\beta(\theta)}} =\mu_{\rm h} \left(\frac{\theta}{\rm arcsec}\right)^{-1}. \label{Icarus_mut}
\end{equation}
The parameters related to the (effective) source size are as 
follows; the average convergence of microlenses $\kappa_{\star} = 0.005$, the mass of the micolenses $M_{\star} = 0.3\ M_{\odot}$, and the physical size of the source $R_{\rm W} = 180\ R_{\odot}$.
In the unit of arcsec, the Einstein radius of a microlens with mass $M_{\star}$ is  
\begin{equation}
    \theta_{\rm Ein} = 1.8 \times 10^{-6} \left(\frac{M_{\star}}{M_{\odot}}\right)^{\frac{1}{2}}\ {\rm arcsec},
\end{equation}
and the source size with $R_{\rm W}$ is 
\begin{equation}
    \sigma_{\rm W} = 2.7 \times 10^{-12} \left(\frac{R_{\rm W}}{R_{\odot}}\right)\ {\rm arcsec}.
\end{equation}
The Icarus is observed at $0.13\ {\rm arcsec}$ away from the macro-critical curve, where the average magnification is approximately $\mu = \mu_{t}\mu_{r} = 300$ from Eq.~\eqref{Icarus_mut}.
This average magnification is about an order of magnitude smaller than the peak magnification of Icarus that is estimated to be a few thousands, which indicates that the observed star is influenced by the microlensing effect to be highly magnified.

In addition to the above parameters, the width of the arc is $w_{\rm arc} = 0.2\ {\rm arcsec}$, and the number density in the lens plane of the sources with the radius between $R_{1}$ and $R_{2}$ is 
\begin{equation}
    n_{\rm source}(R_{1} < R < R_{2}) = \frac{n_{0}}{\mu_{\rm av}} \left\{\left(\frac{R_{\odot}}{R_{1}}\right)^{2} - \left(\frac{R_{\odot}}{R_{2}}\right)^{2}\right\}, \label{Icarus_nsource}
\end{equation}
with $n_{0} = 1.9\times 10^{7}\ {\rm arcsec}^{-2}$.
We use these values to estimate the expected number of high-magnification events in Sec.~\ref{subsec:prediction}.

\subsection{Modeling the total probability distribution function} \label{subsec:total_pdf}
Since our model described in Sec.~\ref{sec:analytical_pdf} only focuses on the high-magnification tail, we describe how we model the total PDF in this subsection.

\citet{2024A&A...687A..81P} numerically show that the PDF around the averaged magnification satisfies the log-normal distribution.
The averaged magnification and its variance are studied in \citet{2021arXiv210412009D} as shown in Sec.~\ref{subsec:review_random_ml}.
Therefore, using the average magnification $\mu_{\rm av}$ and its standard deviation ${\rm Std}[\mu_{\rm av}]$ shown in Eq.~\eqref{mu_std_general}, we model the PDF around the average magnification as
\begin{eqnarray}
    &&\left.\frac{dP}{d\log_{10} \mu}\right|_{\rm middle} \nonumber\\
    =&& \frac{1}{\sqrt{2\pi \sigma_{\log_{10}\mu_{\rm av}}^{2}}} \exp \left( -\frac{(\log_{10}\mu - \log_{10} \mu_{\rm av})^{2}}{2\sigma_{\log_{10}\mu_{\rm av}}^{2}} \right),
\end{eqnarray}
where the standard deviation for the log-normal distribution is
\begin{equation}
    \sigma_{\log_{10} \mu_{\rm av}} = \frac{{\rm Std}[\mu_{\rm av}]}{\mu_{\rm av} \ln 10}.
\end{equation}
Again, the high-magnification tail of the PDF is expressed by 
\begin{eqnarray}
    &&\left.\frac{dP}{d\log_{10} \mu}\right|_{\rm high} = 2\ln 10\ P^{\rm PS}(\mu>\mu_{\rm th}) \nonumber \\
    && \hspace{20mm} \times \frac{1+e^{-1}}{1+\exp\left(\frac{\mu-\mu_{\rm max}}{\mu_{\rm max}}\right)}\left(\frac{\mu}{\mu_{\rm th}}\right)^{-2}.
\end{eqnarray}
Therefore, we combine the above PDFs to obtain the total PDF,
\begin{equation}
    \left.\frac{dP}{d\log_{10} \mu}\right|_{\rm tot} = 
        \begin{cases}
        \left.A_{\rm norm} \frac{dP}{d\log_{10} \mu}\right|_{\rm middle} \ \  (\mu \leq \mu_{0}) \\
        \left.\frac{dP}{d\log_{10} \mu}\right|_{\rm high}\ \ \ \ \ \ \ \ \ \ \ \ (\mu \geq \mu_{0}),
    \end{cases} \label{total_pdf_Anorm}
\end{equation}
where $\mu_{0} > \mu_{\rm av}$ is the matching magnification between the log-normal distribution and the high-magnification tail, and the $A_{\rm norm}$ is the normalization factor such that the total probability is normalized to one.

\begin{figure}
\includegraphics[width=\columnwidth]{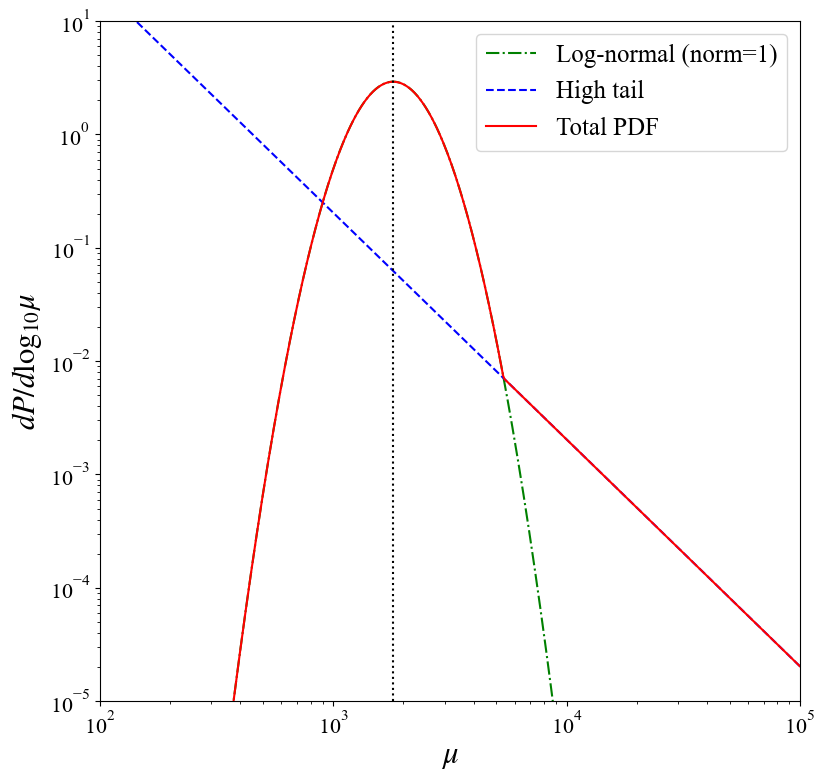}
    \caption{The total PDF of the Icarus-like system at $\theta = 0.039\ {\rm arcsec}$ from the macro-critical curve. 
    The green dash-dot line is the PDF around the average magnification, which is shown in a vertical dotted line.
    The blue dashed line is the high-magnification tail of the PDF.
    The red solid line is the total PDF with the normalization factor $A_{\rm norm} = 0.9987$ introduced in Eq.~\eqref{total_pdf_Anorm}.
    }
    \label{fig:toal_pdf}
\end{figure}

In Fig.~\ref{fig:toal_pdf}, we show the example of the total PDF of the Icarus-like system at $\theta = 0.039\ {\rm arcsec}$ from the macro-critical curve.
From Eq.~\eqref{Icarus_mut}, the corresponding point on the source plane is $\beta = 7.0\times 10^{-5}\ {\rm arcsec}$ away from the caustic.
The average magnification is calculated from Eq.~\eqref{Icarus_muav}.
The variance of the average magnification is calculated from Eq.~\eqref{mu_dis_constant_kg}, which considers the constant background convergence and the shear.
This relation is supposed to be valid locally.
The constant background convergence is calculated from the average magnification above by assuming the shear is the same as the convergence for simplicity.
It is seen that the transition between the middle and the high magnification tail occurs $\mu_{0} \simeq 3\mu_{\rm av}$ in this case.

\subsection{Prediction of high-magnification events} \label{subsec:prediction}
\begin{figure}
\includegraphics[width=\columnwidth]{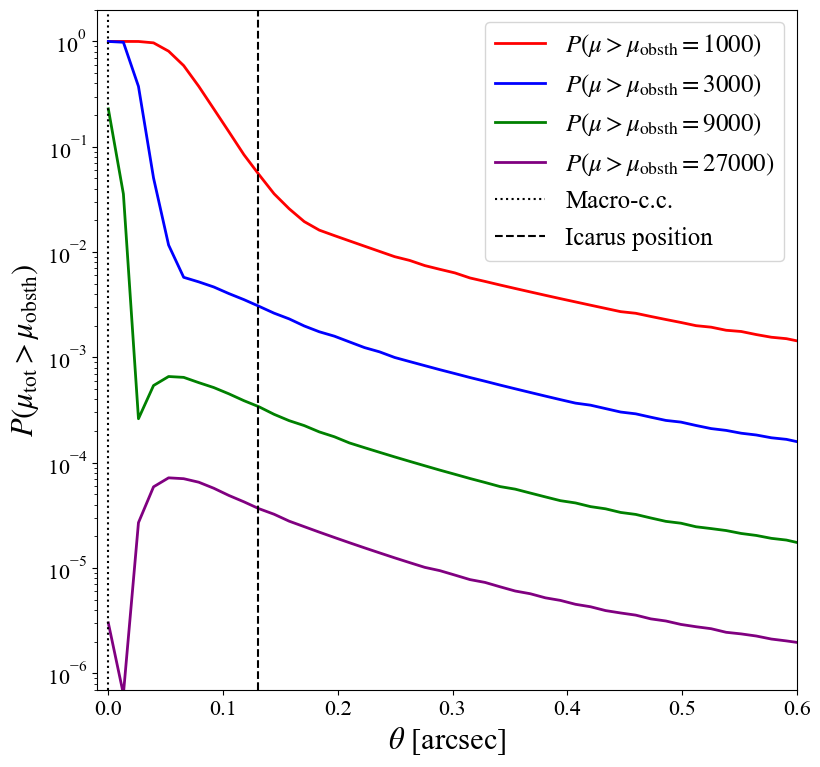}
    \caption{The probability of the total magnification to be larger than the observational threshold $\mu_{\rm obsth}$ as a function of the distance from the macro-critical curve in the Icarus-like system.
    The red, blue, green, and purple lines show the cases with different thresholds $\mu_{\rm obsth} = 1000$, $3000$, $9000$, and $27000$, respectively. 
    The position of the macro-critical curve is shown in a vertical dotted line.
    The vertical black dashed line shows the observed location of the Icarus.
    }
    \label{fig:Icarus_muobsth}
\end{figure}

Since we obtain the total PDF at each point as shown in Sec.~\ref{subsec:total_pdf}, we now calculate the probability that the observed magnification exceeds the threshold $\mu_{\rm obsth}$.
To do so, we first calculate the average magnification and the total PDF, and then calculate the probability $P(r > r_{\rm th})$ by integrating the PDF above the threshold $r_{\rm th} = \mu_{\rm obsth}/\mu_{\rm av}$.
In Fig.~\ref{fig:Icarus_muobsth}, we show the probability to be magnified by a factor of more than the observational threshold $\mu_{\rm obsth} = 1000$, $3000$, $9000$, and $27000$, as a function of the distance from the macro-critical curve, assuming model parameters for the Icarus-like system.
When the thresholds are modest, 1000 and 3000, the probability becomes larger as closer to the macro-critical curve since the average magnification becomes larger. The probability is almost equal to one near the macro-critical curve since the average magnification is sufficiently large compared to the threshold.
When the thresholds are higher, 9000 and 27000, there are local minima of the probability between 0.0 and 0.1 arcsec.
Moreover, we find that the position where the probability becomes maximum is not on the macro-critical curve but around $\theta \simeq 0.05\ {\rm arcsec}$ in the case of $\mu_{\rm obsth} = 27000$.

\begin{figure}
\includegraphics[width=\columnwidth]{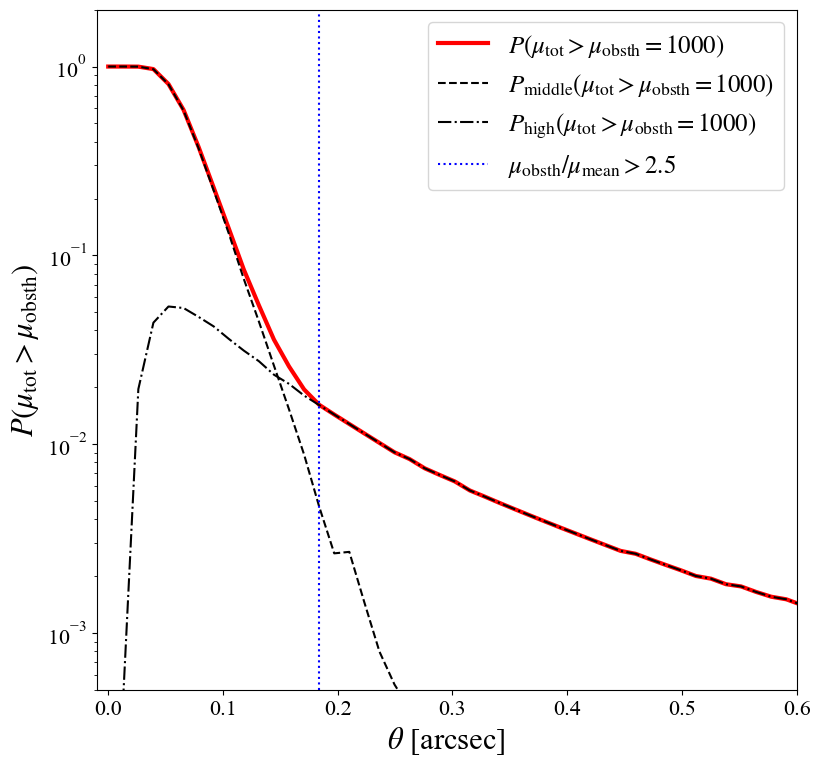}
    \caption{The probability of the total magnification to be larger than the observational threshold $\mu_{\rm obsth} = 1000$.
    Dashed and dash-dot lines show contributions from the middle PDF and the high-magnification tail of the PDF, respectively.
    The vertical blue dotted line shows the approximate position of the transition between them.
    }
    \label{fig:Icarus_P_dev}
\end{figure}

To better understand this behavior, we decompose the integrated total probability into contributions from the middle PDF and the high-magnification tail of the PDF, as shown in Fig.~\ref{fig:Icarus_P_dev}.
For the contribution from the middle PDF around the average magnification, 
when the average magnification is larger than the threshold, $\mu_{\rm av} \gtrsim \mu_{\rm obsth}$, the probability is close to one.
For the contribution from the high-magnification tail of the PDF, 
there is a turnover near the macro-critical curve, below which the probability becomes exponentially suppressed. In our model, this suppression occurs in the nonlinear regime satisfying $\mu_{\rm av} \gtrsim 1/(f_{\star} \kappa_{\rm tot})$.
In contrast, in the linear regime, the probability increases as the distance from the macro-critical curve becomes smaller and the average magnification becomes larger.
This turnover is the same as what we observe in Fig.~\ref{fig:Icarus_muobsth} for high $\mu_{\rm obsth}$. 
The overall behavior of the contribution from the high-magnification tail of the PDF is understood from Eq.~\eqref{P_ps_A0_B0}. Ignoring the source size and approximating the dependence of the average magnification on the distance to $\mu_{\rm av} \propto \theta^{-1}$, the integrated probability satisfies
\begin{eqnarray}
    P(\mu > \mu_{\rm obsth}) &\propto& \mu_{\rm av}^{\frac{5}{2}} \exp(-B_{0} f_{\star} \kappa_{\rm tot} \mu_{\rm av}) \nonumber \\
    &\propto& \theta^{-\frac{5}{2}} \exp(-B_{0} f_{\star} \kappa_{\rm tot} \mu_{\rm h} \mu_{\rm r} \theta^{-1}). 
\end{eqnarray}

Another important feature in Fig.~\ref{fig:Icarus_muobsth} is how different magnification thresholds affect the overall normalization of the probability.
This can be easily understood from the relation $P(\mu > \mu_{\rm obsth}) \propto r_{\rm th}^{-2} \propto \mu_{\rm obsth}^{-2}$.

The position of the observed Icarus is shown in a vertical dashed line in Fig.~\ref{fig:Icarus_muobsth}.
We can see that this position has a sufficient probability to be observed but not the maximum.
We note that Earendel, whose magnification is expected to be $>4000$, is observed exactly on a macro-critical curve,  which is consistent with this result.

\citet{2024arXiv240316989V} model the probability distribution of the observed location.
Within the corrugated band of the micro-critical curves near the macro-critical curve, the probability distribution is estimated as constant and drops proportional to $\theta^{-2}$ outside the band.
Moreover, their model of the probability distribution only changes the overall probability but does not change the functional form with different observational thresholds.
These features are slightly different from our modeling shown above.
Therefore, we expect that by adopting our new analytic model to repeat the analysis in \citet{2024arXiv240316989V}, we can place more robust constraints on the abundance of primordial black holes from observed locations of highly magnified stars. We leave such study for future work.

We now estimate the number of high-magnification events observed in a single snapshot.
The expected number of images that are magnified more than $\mu_{\rm obsth}$ is
\begin{equation}
    N = 2\int^{\theta_{\rm max}}_{\theta_{\rm min}} d\theta\ n_{\rm source} w_{\rm arc} \int_{\mu_{\rm obsth}}^{\infty} \frac{dP}{d\log_{10} \mu} d\log_{10} \mu, \label{Icarus_N}
\end{equation}
where the coefficient 2 indicates that the images are found on both sides of the critical curve.
We set the observation region $\theta_{\rm min} = 0.0\ {\rm arcsec}$ and $\theta_{\rm max} = 1.3\ {\rm arcsec}$.
By substituting Eq.~\eqref{Icarus_nsource} into Eq.~\eqref{Icarus_N} with $R_{1}=140(\mu_{\rm t}/100)^{-1/2}(M/M_{\odot})^{-1/6}\ R_{\odot}$ and $R_{2} = 730\ R_{\odot}$,
\begin{eqnarray}
    N &&= 2\int^{\theta_{\rm max}}_{\theta_{\rm min}} d\theta\ \frac{n_{0} w_{\rm arc}}{\mu_{\rm av}} \left[ 140^{-2} \left(\frac{\mu_{t}}{100}\right) \left(\frac{M_{\star}}{M_{\odot}}\right)^{\frac{1}{3}} - 730^{-2} \right]\nonumber\\
    && \hspace{20mm} \times \int_{\mu_{\rm obsth}}^{\infty} \frac{dP}{d\log_{10} \mu} d\log_{10} \mu.
\end{eqnarray}
From this equation, we calculate the expected number of high-magnified images in a single snapshot.
We obtain $N = 0.073$, $0.028$, and $0.0031$ for $\mu_{\rm obsth} = 1000$, $3000$, and $9000$, respectively.
Considering that only one event is observed in the extensive HST observations for two years and the duration of Icarus to be highly magnified is roughly two weeks, the observed mean number of Icarus-like events in each snapshot is approximated as $N \simeq 1/52 \simeq 0.019$.
The above estimation with our model shows quite a good agreement with the observation, indicating that our model is suitable for future predictions of high-magnification events.

\section{Summary and discussions} \label{sec:discussion_conclusion}
Highly magnified stars such as Icarus and Earendel have recently been observed in galaxy clusters. To properly interpret these observations and to predict the occurrence of similar events in future observations, it is important to understand how the high-magnification tail of the PDF is affected by various parameters, such as the mass fraction of microlenses $f_{\star}$, originating from ICL stars for instance, and the average magnification $\mu_{\rm av}$.
To achieve this goal, we propose a new physical model of the high-magnification tail with the help of ray-tracing simulations conducted by {\sc CCtrain} and {\sc Gerlumph}. 

Our analytic model assumes that the probability is proportional to the independent number of micro-critical curves whose nearest neighbor inter-microlens distance is larger than the size of its micro-critical curve.
We first consider the case with a point source, where there is no maximum magnification, and show that the parameter dependencies on the PDF are encapsulated in the form of $X = f_{\star} \kappa_{\rm tot} \mu_{\rm av}$.
In the linear regime, $X \lesssim 1$, the number of independent micro-critical curves is almost the same as that of microlenses.
Therefore, the (integrated) probability scales linearly with $X$.
In the nonlinear regime, where $X$ is larger than one, the micro-critical curves merge and the independent number of the microlenses decreases exponentially, resulting in the probability being exponentially suppressed.
The validity of our model in the case of a point source is confirmed by using the simulation data of {\sc CCtrain}.

We then consider the case with a finite source size.
Due to the averaging effect within the source size, there is a maximum magnification and the high magnification tail of the PDF is suppressed above the maximum magnification.
We introduce the suppression factor $S(r;r_{\rm max})$ to consider this effect to find that the sigmoid function is suitable to explain the simulation data by {\sc Gerlumph}.
We note that the resolution on the source plane of {\sc Gerlumph} can be regarded as an effective source size.
Combining all the results, we obtain our analytic model as described in Eq.~\eqref{eq:model_summary_1} with Eqs.~\eqref{P_ps_A0_B0} and \eqref{rmax_lin_nonlinear}.

Our analytic model allows us to investigate parameter degeneracies in the high-magnification tail of the PDF. For instance, there are two degeneracies we can immediately infer from our model.
The first one is the combination of the parameters $f_{\star} \kappa_{\rm tot} \mu_{\rm av}$, which is crucial to understanding the behavior of the high-magnification tail.
It indicates that the fraction of stars degenerate with the smooth lens mass distribution.
This degeneracy could be alleviated with detailed mass modeling using multiple strong lens images.
The second one is the combination of the parameters $\theta_{\rm Ein}/(\sqrt{\mu_{\rm av}}\sigma_{\rm W})$ existing in the maximum magnification $r_{\rm max}$.
It indicates that the source size (or source mass) degenerates with the mass of microlenses.

For the application of our model, we predict the expected number of highly-magnified events in a single snapshot and the probability distribution of their observed locations, assuming model parameters that are relevant to the case of Icarus.
We assume that the PDF around the average magnification is described by a log-normal distribution to obtain a model of the total PDF covering all the magnification values. 
At each point near the macro-critical curve (caustic), we calculate the average magnification and obtain the probability that the magnification exceeds the observational threshold.
As closer to the macro-critical curve, the probability tends to become larger since the average magnification becomes larger, although there is a non-trivial behavior in some cases.
The expected number of images is calculated as $\mathcal{O}(10^{-2})$ when the observational threshold is set to $\mu_{\rm obsth} = \mathcal{O}(10^{3})$.
We find that our model prediction agrees well with the HST observations in which a single event continuing for roughly two weeks is observed in a two-year survey, confirming the validity of our analytic model as a powerful tool to predict the future occurrence of highly magnified stars.

One limitation of our model is that we only consider microlenses with the same mass.
For example, the mass function of ICL stars and stellar black holes are affected by the initial mass function, stellar evolution history, and core-collapsed physics, and they do not have the same mass \citep{2002RvMP...74.1015W, 2012ApJ...749...91F, 2015MNRAS.451.4086S}.
Primordial black holes (PBHs) are also candidates to contribute to the microlenses, which are considered to have a wide range of masses \citep{2020ARNPS..70..355C, 2021JPhG...48d3001G}.
While we do not consider these cases with a mass spectrum, we expect that the PDF would not be changed in the linear regime since the total area of each magnification bin is unchanged, which is easily understood from the fact that the Einstein radius scales with the mass of the microlenses $M_{\star}$ as $\theta_{\rm Ein} \propto M_{\star}^{1/2}$ and the number density of the stars scales as $n_{\star} \propto M_{\star}^{-1}$ and as a result the total area that is proportional to $n_{\star} \theta_{\rm Ein}^{2}$ is unchanged.
In the nonlinear regime, however, we expect that the PDF could be changed in a complex manner.
For example, when the population of microlenses follows a power-law distribution and there are numerous small objects, we expect that the high-magnification tail of the PDF would be suppressed since the number of independent micro-critical curves originating from microlenses with larger mass would be smaller.
Inversely, these small objects would create smaller micro-critical curves and some of them would be independent, which might amplify the amplitude of the PDF.
However, since the typical magnification is proportional to the size of the micro-critical curves as shown in Eq.~\eqref{mag_near_crit_w_bg}, the amplification of the PDF would occur only in the middle region of the high-magnification tail, rather than the highest magnification tail that would be suppressed as stated above.
We leave the extension of our analytic model to consider a mass spectrum for future study, from which we expect more robust predictions of high-magnification events and constraining the population of microlenses, {\it{e.g.}}, the PBH abundance, can be achieved.

Another simplified assumption in our model is the spatial distribution of microlenses.
We assume that the microlenses are uniformly distributed randomly.
The spatial scale required for this assumption to hold and for our model to be valid is several times the size of the microcritical curve.
However, some microlenses may be binary systems.
In that case, the distribution of the microlenses could be non-uniform below the size of the micro-critical curves, and the number of independent micro-critical curves could be changed.
Moreover, since the shape and the size of the micro-critical curves change a lot according to masses and the relative distances of the binary systems, the PDF may be changed from what our model predicts.
The effect of the spatial distribution of microlenses on the magnification PDF might be larger in the linear regime than in the nonlinear regime.

Using our analytic model, possibly with more realistic extensions, we can predict the event rate of high-magnification events and their observed positions, as shown in this paper in the Icarus-like system.
Moreover, we can use our model to constrain the mass fraction of microlenses from the observed event rate and positions.
By comparing the mass fraction obtained from the above analysis and that obtained directly from the photometric observation of ICL by assuming the stellar-to-light ratio, we can constrain the unresolved microlenses such as PBH.
As such, our analytic model of the magnification PDF combined with the statistical observational data is a powerful tool for estimating the parameter space statistically.
We expect that our study is an important step toward uncovering the stellar evolution history as well as the nature of dark matter.

\begin{acknowledgments}
We thank the anonymous referee for the useful comments.
We thank Georgios Vernardos, Jordi Miralda-Escud{\'e}, Jos{\'e} Mar{\'i}a Diego, Tom Broadhurst, and Jos{\'e} Mar{\'i}a Palencia for useful feedback during this project.
This work was supported in part by JSPS KAKENHI Grant Numbers JP22J21440, JP22H01260, and JP20H05856.
H. K. thanks the hospitality of INAF OAS Bologna where part of this work was carried out and useful discussion took place.
It is supported by JSPS KAKENHI Grant Numbers JP22K21349.
\end{acknowledgments}

\bibliography{ref}

\appendix
\section{Lens potential} \label{app:lens_potential}
To study lensing properties near the critical curve, it is useful to set the origins of the lens and the source planes on a critical curve and on a caustic, respectively, and expand the lens potential around the origin.
The expression of the Taylor expansion is written as
\begin{eqnarray}
    \psi(\boldsymbol{\theta}) = && \psi(\boldsymbol{0}) + (\psi_{,1}(\boldsymbol{0}) \theta_{1} + \psi_{,2}(\boldsymbol{0}) \theta_{2}) \nonumber \\
     &&+ \frac{1}{2} (\psi_{,11}(\boldsymbol{0})\theta_{1}^{2} + 2\psi_{,12}(\boldsymbol{0})\theta_{1}\theta_{2} + \psi_{,22}(\boldsymbol{0})\theta_{2}^{2}) \nonumber \\
     &&+ \frac{1}{6} (\psi_{,111}(\boldsymbol{0})\theta_{1}^{3} + 3\psi_{,112}(\boldsymbol{0})\theta_{1}^{2}\theta_{2} \nonumber \\
     && \hspace{10mm} + 3\psi_{,122} (\boldsymbol{0}) \theta_{1}\theta_{2}^{2} + 
     \psi_{,222}(\boldsymbol{0}) \theta_{2}^{3}).
\end{eqnarray}
Here we can ignore the zeroth and first order terms since these yield the constant shifts of the potential and the deflection. 
The convergence of the origin is denoted by $\kappa_{0}$, which can be written with the lens potential as
\begin{equation}
    \kappa_{0} = \frac{1}{2} (\psi_{,11}(\boldsymbol{0}) + \psi_{,22}(\boldsymbol{0})).
\end{equation}
Then, we can write 
\begin{eqnarray}
    &&\psi_{,11}(\boldsymbol{0}) = \kappa_{0} + (1-\kappa_{0})\cos \omega, \\
    &&\psi_{,22}(\boldsymbol{0}) = \kappa_{0} - (1-\kappa_{0})\cos \omega,
\end{eqnarray}
where $\omega$ is an arbitrary constant.
Since the origin is on the critical curve, the determinant of the lens matrix is zero, $\det A(\boldsymbol{0}) = 0$.
From this condition, we can obtain
\begin{equation}
    \psi_{,12}(\boldsymbol{0}) = -(1-\kappa_{0})\sin \omega.
\end{equation}
For simplicity, we can choose $\omega = 0$.
Additionally, we consider the completely orthogonal system, {\it{i.e.}}, a train of the micro-images is perpendicular to the critical curve.
By setting the critical curve to coincide with $\theta_{2}$ axis, $\det A((0,\theta_{2})) = 0$, we can obtain $\psi_{112}(\boldsymbol{0}) = \psi_{122}(\boldsymbol{0}) = 0$.
From the condition that the two images produced by the same source are symmetrical about the critical curve, we obtain $\psi_{222}(\boldsymbol{0}) = 0$.
Finally, by setting $\psi_{111}(\boldsymbol{0})=-\epsilon$, we can obtain the lens potential of Eq.~\eqref{lens_pot_near_cc}.

The quantity $\epsilon$ is proportional to the curvature of the critical curve. 
To see this, let us consider the point mass lens case. 
The lens potential is
\begin{equation}
    \psi(\bar{\theta}) = \theta_{\rm Ein}^{2} \ln(\bar{\theta}),
\end{equation}
where the origin is the center of the point mass lens and $\bar{\theta}$ denotes the radial distance from the origin.
By expanding the lens potential around a point on the critical curve in the radial direction, $\bar{\theta} = \theta_{\rm Ein} + \theta$, we can express the third order term as
\begin{equation}
    \psi_{\rm 3rd}(\theta) \simeq \frac{1}{3 \theta_{\rm Ein}} \theta^{3}.
\end{equation}
Therefore, we can see that $\epsilon \propto \theta_{\rm Ein}^{-1}$.

\end{document}